\documentclass[aps,pre,floatfix,twocolumn,footinbib]{revtex4}
\usepackage{epsfig}
\usepackage{graphicx}
\usepackage{amsmath}
\usepackage{appendix}
\usepackage{mathrsfs}  
\newcommand{\bx}{\mathbf{x}}

\newcommand{\bk}{\mathbf{k}}
\newcommand{\bq}{\mathbf{q}}

\newcommand{\bT}{\mathbf{T}}

\newcommand{\hpsi}{\hat{\psi}}

\newcommand{\om}{\omega}
\newcommand{\omm}{\omega_\mu}
\newcommand{\sd}{\tilde{d}}

\newcommand{\be}{\begin{equation}} 
\newcommand{\ee}{\end{equation}} 
\newcommand{\bea}{\begin{eqnarray}} 
\newcommand{\eea}{\end{eqnarray}} 
\newcommand{\bc}{\begin{center}} 
\newcommand{\ec}{\end{center}} 

\begin{document}

\title{Spectral Renormalization Group for the Gaussian model and $\psi^4$ theory on non-spatial networks}
\author{Asl{\i} Tuncer$^{1,2}$ and Ay\c{s}e Erzan$^1$}
\affiliation{$^1$Department of Physics, Istanbul Technical University, Maslak 34469, Istanbul, Turkey\\
$^2$Department of Physics, I\c{s}\i k University,  \c{S}ile 34980, Istanbul, Turkey
 }

\begin{abstract}

We implement the spectral renormalization group on different deterministic non-spatial networks without translational invariance. We calculate the thermodynamic critical exponents for the Gaussian model on the Cayley tree and the diamond lattice, and find that they are functions of the spectral dimension, $\sd$. The results are shown to be consistent with those from exact summation and finite size scaling approaches.  At  $\sd=2$, the lower critical dimension for the Ising universality class, the Gaussian fixed point is stable with respect to a $\psi^4$ perturbation up to second order. However,  on generalized diamond lattices, non-Gaussian fixed points arise  for $2<\sd<4$.


\end{abstract}
\date{\today}

\maketitle

\section{Introduction}

Both static and dynamical  phenomena on networks, which typically lack translational invariance and may not be naturally embedded in a metric space, have been the subject of intense study over the last decade and a half ~\cite{BA_book,Vespignani_largescale,Vespignani_Dynamical}. Phase transitions and critical phenomena on networks have also received a lot of attention ~\cite{BA,dorogovtsev1}. 

To date,  a unified approach to the theory of critical phenomena on arbitrary networks,   analogous  to the renormalization group theory  developed by Wilson and Kogut~\cite{Wilson1,Wilson2,Wilson_F,Wilson3} on  periodic networks, is still  lacking. 
The outstanding achievement of the renormalization group (RG) theory of critical phenomena  was to explain the experimentally observed phenomenon of ``universality'' and introduce such concepts as the ``relevance'' or ``irrelevance'' of different types of interactions, upper and lower critical dimensions  and  the elucidation of the roles of the dimensionality of space and of the order parameter.  

Dorogovsev {\it et al.}~\cite{dorogovtsev1} have shown how the critical behavior of scale free graphs depends on the scaling exponent $\gamma$ of the degree distribution, and Bradde {\it et al.}~\cite{Bianconi2010} have derived a Ginzburg criterion in terms of an effective spectral dimension for spatial scale free networks. 
Various ``real space renormalization group'' (RSRG)~\cite{Leeuwen,Burkhardt,newKadanoff}  methods have been proposed for arbitrary networks~\cite{Havlin}, but even when they are exact, they rarely reveal universal properties of critical phenomena on non-spatial networks in terms of their spectral and topological properties, and there is still room for improving our understanding. 

In a recent publication~\cite{erzan} we have proposed a spectral  renormalization group (SRG) scheme modelled on  the ``momentum shell" renormalization group a l\`a Wilson~\cite{Wilson3}.  We expand the fluctuations of the order parameter in terms of the eigenvectors of the graph Laplacian~\cite{Chung}  in a generalized Fourier transform, partly surmounting the difficulty posed by non-translationally invariant lattices. Elimination of the large eigenvalue fluctuations and rescaling of the effective Hamiltonian then yield, in the same spirit as in the Wilson renormalization group, the rescaling factors for the coupling constants, which can then be related to the critical exponents. 

On  non-translationally invariant, non-spatial networks, the eigenvalues of the graph Laplacian  do not have an obvious interpretation in terms of lattice momenta and an isotropic, translationally invariant correlation length is not available.   Therefore the exponent of the RG eigenvalue in the temperature-like direction under length- rescaling cannot be naively interpreted in terms of an inverse correlation length exponent. We have to develop RG schemes which do not involve length-like concepts.  
 
In this paper, we explicitly  implement the SRG~\cite{erzan}  on two non-spatial networks which lack translational invariance, namely, the Cayley tree and the diamond lattice \cite{diamond} for Gaussian model~\cite{Berlin,Huang,Goldenfeld}. Then we include a quartic interaction term, which on periodic lattices is known to carry  the ``trivial'' theory into the Ising universality class.\cite{Wilson3,Amit} and we investigate the renormalization behavior of the interacting theory within a perturbation expansion up to the second order in the coupling constant and the deviation from the critical temperature.   

We find that for the Gaussian theory, the critical exponents depend only on the spectral dimension of the lattice, i.e.,  the  scaling behavior of the small eigenvalue region of the  Laplace spectrum.   However,  the interacting theory depends sensitively on the symmetry properties of the lattice, via the eigenvectors of the graph Laplacian, which enter the calculations of the four-vertex. 

Within second order perturbation theory we find that the Gaussian fixed point is stable at $\sd$.  Extending our calculations to a series of generalized diamond lattices with  higher spectral dimensions, we establish the existence of non-trivial fixed point of the SRG for $2\,< \,\sd \,<\, 4$. 

In Section~II, we define  the spectral renormalization group for the Gaussian model on a generic network.    In Section~III, we implement this scenario on the Cayley tree and the diamond (hierarchical) lattice. We compute the specific heat and magnetic field exponents. 
For comparison, exact enumeration and finite size scaling results are presented in Section IV.     
 In Section~V we include $\psi^4$ interactions.  
In Section~VI we provide a discussion and conclusions.

\section{The Spectral Renormalization Group for the Gaussian Model} 

The effective Ginzburg-Landau ``Lagrangian" for a  scalar order parameter $\psi(\bx)$ is given by,
\be
\begin{split}
H= \int_V d {\bx} \{\frac{1}{2} \left[ r_0 \psi^2(\bx) - \psi(\bx)\nabla^2\psi(\bx) \right]+ \\
 v_0 \psi^4(\bx) - h \psi(\bx)\}\;\;.\label{eq:landau}
\end{split}
\ee
where the integral is over the volume of the system, $r_0$ is proportional to the reduced temperature $t=(T-T_c)/T_c$ and $h$ to the magnetic field. We will assume that $H$ is expressed in units of the thermal energy  $k_B T$ where $k_B$ is the Boltzmann constant. The Gaussian model~\cite{Berlin} is equivalent to omitting the fourth order coupling term in Eq.~(\ref{eq:landau}).  This model is defined only for temperatures above the critical temperature, i.e., for $r_0 >0$.  Nevertheless one may formally compute the exponent $\delta$.

For a continuous field $\psi(i)\in (-\infty, \infty)$, living on the nodes of an arbitrary network, the Gaussian model can be written as 

\begin{equation}\label{eq:GH}
H_0=\frac{1}{2}\sum_{ij}^N\psi(i)[r_0\delta_{ij}+L_{ij}]\psi(j)-h\sum_i \psi(i) \;\;.
\end{equation}
The usual Laplace operator appearing in the Ginzburg-Landau expansion  has been replaced by (minus)
the graph Laplacian~\cite{Chung}, with the matrix elements,
\begin{equation} \label{eq:AELaplace}
L_{ij}=d_i\delta_{ij}-A_{ij}\;\;,
\end{equation}
where ${\bf A}$ is the adjacency matrix of the network 
and $d_i$ is the degree of the $i$th node. The expression in 
Eq.~(\ref{eq:GH}) is now very general, applicable to arbitrary networks, with only the requirement that the matrix ${\bf A}$ be symmetric, so that its eigenvalues are real. 

Expanding the field $\psi(i)$ in 
terms of eigenvectors $\mathbf{u}_\mu$ of the Laplace operator, 
$\psi(i)=N^{-1/2}\sum_\mu \hpsi_\mu u_\mu(i)$,
 the Hamiltonian is obtained in diagonal form, 
\begin{equation}\label{eq:H}
H_0=\frac{1}{2}\sum_{\mu=1}^N[r_0+\omega_\mu]\hpsi_\mu^2- h \hpsi_1\;\;.
\end{equation}
Here $\omega_\mu$ are the eigenvalues of 
${\bf L}$. The eigenvalues are ordered so that $\om_1\le \om_2\ldots \le \om_N$,  with $\omega_1=0$.  We will assume the network to be connected so that $\om_2>0$ for finite $N$.  

For this system, the partition function is immediately obtained from 
\be 
\label{eq:Z0}
Z_0 = \int_{-\infty}^\infty \prod_\mu d\hpsi_\mu e^{-H_0}\;\;,
\ee
and the free energy is given, up to constant terms, by 
\be \label{eq:FE}
F_0=\frac{1}{2}\sum_{\mu=1}^N \ln(r_0+ \omm) -\frac{h^2}{2r_0}\;\;.
\ee
(Henceforth we will drop the external field term unless we are directly dealing  with it.)

Note that in Eq.(\ref{eq:H}) there is a difficulty in going over from a sum (over $\mu = 1,\ldots,N$) to an integral over the eigenvalues.   In general the eigenvectors $\mathbf{u}_\mu$, and consequently the $\hpsi_\mu$, do not possess, e.g.,  the rotational symmetries valid on periodic lattices.  Therefore in general it is {\em not} justified to try to extract the renormalization factors by rewriting the Hamiltonian as, 
\begin{equation}\label{eq:Hbeta}
  \frac{1}{2}\int_{0}^\Omega d\om \rho(\om){[r_0+\omega]\hpsi_\om^2}\;\;,
\end{equation}
where $\Omega$ is the largest eigenvalue.
On the other hand, after the Gaussian integrals have been carried out, this difficulty is not there for the free energy (or its derivatives, such as the specific heat or the two-point correlation function~\cite{Hattori})  and one may formally write,
\be \label{eq:FEbeta}
F_0=\frac{1}{2}\int_{0}^\Omega d\om \rho(\om) \ln(r_0+ \om)  \;\;.
\ee

We would now like to show how we can implement field theoretic renormalization group ideas on a system which, besides not having an a priori known spectral density,  is not embedded in a metric space, i.e., there is no concept of length.   

We have two possible strategies for eliminating the large $\omega$ fluctuations from the partition function and computing the renormalization factors. In the absence of a ``length like" quantity, the first method which comes to mind is to truncate the number of {\it modes}, $N$, by a constant factor, successively integrating out those with the the largest eigenvalues.
The second method consists of scaling the largest eigenvalue, $\Omega$ by a constant $B$, in analogy with the usual renormalization group a l\`a Wilson~\cite{Wilson1,Wilson2,Wilson3,Goldenfeld}. These two strategies are implemented below, and give identical results for the Gaussian model.

%
%
It should be noted that on these non-spatial lattices, eigenvectors with the same symmetry properties may have widely differing eigenvalues. (We illustrate this in Table~\ref{t:eigenvectors} in  the Appendix, for $N=13$ on the Cayley tree. A similar situation also holds for the diamond lattice.) 
Eliminating those fluctuations associated with the high-$\om$ side of the spectrum makes sense  in terms of eliminating the higher-energy modes but cannot be naively interpreted as eliminating the ``small wavelength'' or ``high frequency'' fluctuations.   On the other hand, if we interpret the successive iterations in the construction of trees or  hierarchical lattices as a fine-graining operation~\cite{GriffithsandKaufman}, the increasing localization of the eigenvectors on the most recently added nodes may be thought of as  greater articulation on smaller scales.

\subsection{Scaling the number of modes}
Since the eigenvalues are numbered in increasing order by convention, we keep the first $N/B$ eigenvalues in the effective Hamiltonian and integrate out the rest. 
Picking the scale factor $B$ in keeping with the overall symmetries of the system is convenient; if no such obvious scale symmetry is available, $B=N/n$, with $n$ integer, eliminates spurious points from the scaling plots. Defining   the  cutoff $\mu_B = N/B $, the truncated Hamiltonian is, 
\begin{equation}
H_0^<=\frac{1}{2}\sum_{\mu=1}^{\mu_B} [r_0+\omega_\mu]  \hpsi_\mu^2 - h \hpsi_1 \;\;.
\end{equation}
Restoring the Hamiltonian to its full range calls for rescaling factors to be inserted, viz.,  
\begin{equation}\label{eq:Hprime}
H_0^\prime =\frac{1}{2}\sum_{\mu=1}^N [r_0 B^{-\phi_1}+B^{-\phi_1-\phi_2} \om_\mu] 
(\hpsi_\mu^\prime)^2 - h^\prime \hpsi_1^\prime \;\;.
\end{equation}
where $\hpsi^\prime = z \hpsi$, and $z$ is the so called ``wave function renormalization."~\cite{Wilson3,Goldenfeld}

We define the rescaling factors $\sigma^{\rm V}_1$ and $\sigma^{\rm V}_2$ as 
\begin{equation}\label{eq:sigma1}
\ \sigma^{\rm V}_1(B)\,\equiv\,\frac{\sum_{\mu=1}^N 1}{\sum_{\mu=1}^{\mu_B} 1}=B^{\phi_1}\;\;
\end{equation}
 and
\begin{equation}\label{eq:sigma2}
\ \sigma^{\rm V}_1(B)\sigma^{\rm V}_2(B)\equiv\frac{\sum_{\mu=1}^N \omega} {\sum_{\mu=1}^{\mu_B} \omega}=B^{\phi_1+\phi_2}.
\end{equation}
where clearly  $\phi_1=1$ , while  $\sum_{\mu=1}^N \omega = N \overline{\omega}$.

To find $z$ we require the coefficient of the Laplace term to remain fixed, and get  $(\sigma^{\rm V}_1 \sigma^{\rm V}_2)^{-1} z^2=1$, which yields
\be 
z= (\sigma^{\rm V}_1\sigma^{\rm V}_2)^{1/2} = B^{(\phi_1+\phi_2)/2}\;\;.
\label{eq:z}
\ee
The renormalization of the reduced temperature is then given by,
\begin{equation}\label{eq:r0}
\ r'=(\sigma_1^V)^{-1}z^2 r_0 = \sigma_2^V r_0= B^{\phi_2} r_0 \;\;.
\end{equation}
The external field term  rescales as $h' \hpsi_1^\prime =h z \hpsi_1$, so that 
\be
h^\prime=zh \;\;.\label{eq:h}
\ee

On this non-translationally invariant network, in the absence of a metric, there is no obvious interpretation of  $\phi_2$ in terms of a correlation length exponent.  Therefore we use the Kadanoff scaling relations in order to express other critical exponents in terms of these renormalization group eigenvalues.  
The Kadanoff scaling relations for the renormalized free energy per mode are
\be f(t,h) = B^{-1}  f^\prime (B^{Y_t^V} t, B^{Y_h^V} h) \;\;.\label{eq:kadanoff}
\ee
From Eqs.~(\ref{eq:r0},\ref{eq:h}), we find $Y_t^V=\phi_2$ and $Y_h^V = (1+\phi_2)/2$.  Finally  
\be f(t,0) \sim t^{2-\alpha}\;\;,\label{eq:AE_scale}
\ee 
yields the specific heat critical exponent $\alpha$,   since  $c_h\sim \partial^2 f(t,0)/\partial t^2  \sim t^{-\alpha}$, 
 \begin{equation}\label{eq:alfafi2}
\ \alpha=2-\frac{1}{\phi_2}\;\;. 
\end{equation}

Setting $t=0$ in Eq.(\ref{eq:kadanoff}), we similarly obtain the magnetic field critical exponent  on the critical isotherm, $h\sim m^\delta$,  with $m$ being  the magnetization per spin.
 \begin{equation}\label{eq:delta1}
\ \delta=\frac{1+\phi_2}{1-\phi_2}.
\end{equation}
\subsection{Scaling the maximum eigenvalue}
 
An alternative strategy  for eliminating degrees of freedom with large $\om$ is to eliminate all degrees of freedom with $\om \ge \Omega/B$, where $B$ is again an arbitrary scale factor. In this case, we define the scaling factors $\sigma^{\Omega}_1$ and $\sigma^{\Omega}_2$ as 
\begin{equation}
\ \sigma^{\Omega}_1(B)\,\equiv\,\frac{N}{\sum_{\mu=1}^{\mu_B} 1}=B^{p_1}\;\;,
\label{eq:sigma1O}
\end{equation}
where
\be \mu_B = \sup \{ \mu \in [1,N] : \omm < \Omega/B \}, 
\label{muBOmega}
\ee
and  $p_1$ is now a non-trivial scaling exponent, with $N/ N^\prime =B^{p_1}$.   
We also have,
\begin{equation}
\ \sigma^{\Omega}_1(B)\sigma^{\Omega}_2 (B)\,\equiv \,\frac{N \overline{\om}} {\sum_{\mu=1}^{\mu_B} \omega}=B^{p_1+p_2}\;\;.
\label{eq:sigma2O}
\end{equation}
Using $\sigma^{\Omega}_1$ and $\sigma^{\Omega}_2$ to rescale the truncated Hamiltonian, one can derive in a way completely analogous to Eqs.~(\ref{eq:Hprime}-\ref{eq:h}), that 
\be z=B^{(p_1+p_2)/2}
\label{eq:zOmega}
\ee  
The recursion relation for the reduced temperature is given by, $r^\prime = \sigma_2^\Omega r_0$.
From simple power counting one has $p_2 =1$.
 
Now taking 
\be 
f =B^{-p_1} f^\prime (t^\prime, h^\prime)\;\;,
\label{eq:fOmega}
\ee
with $t^\prime = B^{ Y_t^\Omega} t$ and  $h^\prime =B^{Y_h^\Omega} h$, one finds that $ Y_t^\Omega = p_2=1$ and $ Y_h^\Omega = (1+p_1)/2$. From  Eq.~(\ref{eq:AE_scale}) one gets,  
\be \alpha = 2-\frac {p_1} {p_2} = 2-p_1 \;\;,\label{eq:alfa_p}
\ee
and similarly,
\be
\delta=\frac {p_1+p_2}{ p_1-p_2} = \frac {p_1+1}{ p_1-1} \;\;.\label{eq:delta_p}
\ee

\begin{figure}[ht]
\vspace{1pt} 
\includegraphics[width=0.5\textwidth]{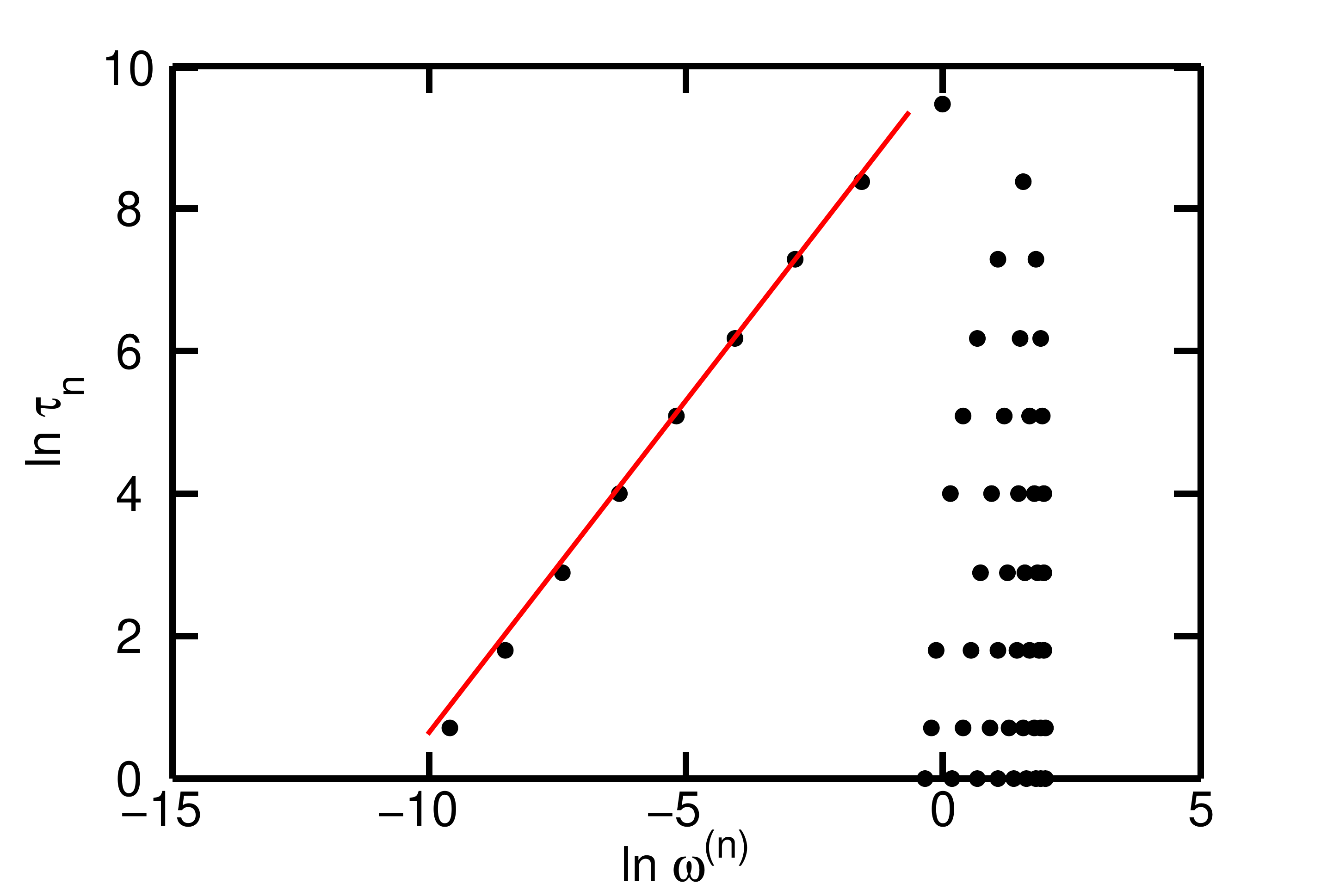}
\caption{(Color online) The degeneracies, $\tau_n$, of the distinct eigenvalues $\omega^{(n)}$ for the Cayley tree (with branching number $b=3$), drawn for $r=9$ generations. The smallest nonzero eigenvalue tends to zero as $b^{-r}$. We do not display $\omega_1=0$. The degeneracies fall on a straight line with unit slope (red in color) in this log-log plot. Nevertheless, the spectral density is zero almost everywhere within the $\om < 1$ domain, and the exponent $\beta$, defined via $\rho(\om)\sim \om^\beta$ for small $\om$ is equal to zero. See text.}\label{fig:Cayley_rho} 
\end{figure}

It is easy to show (from Eqs.(\ref{eq:sigma2},\ref{eq:sigma2O}) that if the spectral density exhibits a power law behaviour, with $\rho(\om)\sim \om^\beta$ for small $\om$, then   the nontrivial exponents $p_1$ and $\phi_2$,  within the context, respectively, of scaling the maximum eigenvalue or the number of modes,  are related to $\beta$ via 
\be \phi_2 = 1/(1+\beta)  \;\;\;\; p_1= 1+\beta \;\;.
\label{p_beta}
\ee
From $\beta\ge 0$, we are ensured that  $\phi_2\le 1$ and  $p_1\ge 1$, and finally, 
\be 
\alpha =1-\beta \;\;\;\; \delta = (2+\beta)/\beta\;\;.
\label{exp_beta}
\ee
In terms of the spectral dimension~\cite{Bianconi2010}  $\sd\equiv 2(1+\beta)$ one gets.
\be
\alpha =\frac{4-\sd}{2} \;\;\;\; \delta =\frac{\sd+2}{\sd-2}\;\;,
\ee
so that for the Gaussian model, the exponents depend solely on the spectral dimension.  A comparison with the exactly known Gaussian exponents in spatial dimension  $d$ ~\cite{Huang,Goldenfeld} shows that  here the spectral dimension has taken on the role of the spatial dimension.

\section{Spectral RG for some Deterministic Networks}
In this section we present numerical and semi-analytical results for the spectral renormalization of the Gaussian model on the Cayley tree and the diamond lattice.  In the Appendix, the analogous computations for the  square and cubic lattices are presented for comparison. 

\subsection{Gaussian model on the Cayley tree}
An inspection of Fig.~\ref{fig:Cayley_rho} shows that the spectral density of the graph Laplacian for the Cayley tree can be written as
\be \rho(\om) = \sum_{n=1}^{r} \tau_n \delta(\om - \om^{(n)})\;\;,
\ee 
%
where 
$ \om^{(n)}$ and $\tau_n=\tau(\om^{(n)})\propto \om^{(n)}$
 are the $n=1, 2,\ldots$th distinct eigenvalues and their degeneracies in the  interval $ 0 <\om < \omega^*$, where $\omega^*$ is the value at which $\tau(\omega)$ is maximum. 

 We see that, for branching number $b$ and $n>1$, 
\be \tau_n=b^{n-2}(b-1) \label{tau}
\ee
and  
\be
\omega^{(n)}\simeq a_n b^{-(r-n+2)}\;\;,\label{omega}
\ee
 where clearly 
\be 
\tau_n \propto [ \omega^{(n)}]^\xi\;\;.
\label{eq:tau-omega}
\ee
In fact we find $\xi = 1$.  The coefficients $a_n$  
tend to a constant, with  $a_{n+1}/a_n \sim 1+ {c_1} \exp(c_2 n)$ for $r-n \gg 1$,  with $c_1$ of the order of $e^{-(r-2)}\ll 1$ and $c_2\sim O(1)$. Since we are interested in the small $\omega$ region of the spectrum we will henceforth treat the $a_n$ as constants. 
The number of eigenvalues within the interval $\om<1$ is $\simeq b^{r-1}\sim N_r/b$.  

To find the spectral dimension, i.e.,  the scaling form of the spectral density, let us consider going  from the discrete sum over $n$  to a continuous integral. 
Defining the continuous variable $x$ via  $\omega^{(n)} \propto   \exp (x\,\ln b)$  and using Eq.~(\ref{eq:tau-omega}),
\be
\sum_n \tau_n \propto  \int dx \exp( \xi x\; \ln b ) \;\;.
\ee
Making the change of variables  $\om(x) = \exp (x\,\ln b)$, we get $d\om=\ln b \exp(x\; \ln b ) dx$ or
\be dx=\frac{ d\om}{ \om \;\ln b}\;\;.\ee
Thus 
\be\sum_n \tau_n  \to   \int \frac{d\om\; \om^\xi} {\om\; \ln b} = \frac{ 1}{\ln b} \int  \om^{\xi-1} d\om\;\;. 
\label{beta}
\ee
This yields $\rho(\omega)\sim \om^\beta$ with $\beta ={\xi-1}$, and since we have found  $\xi=1$, we see that   $\beta = 0$ in the region $\om <1$. Notice that the spectral density (and therefore also the spectral dimension) do not depend on the branching number $b$.

It is straightforward to directly compute the renormalization group eigenvalues from a knowledge of the structure of the discrete eigenvalue spectrum.
Choosing $B=b^k \equiv B_k$,  we have, for the rescaling factors, 
 \begin{equation}\label{eq:cayleyfi1}
\ {\sigma^{\rm V}_1(B_k)} = \frac{N } { \sum_{n=0}^{r-1-k} \tau_n} = B^{\phi_1}\;\;,
\end{equation}
and
 \begin{equation}\label{eq:cayleyfi2}
\  \sigma^{\rm V}_1(B_k)\sigma_2^{\rm V}(B_k)= \;\frac{N \overline{\om}}{ \sum_{n=0}^{r-1-k} \tau_n \omega^{(n)}} \propto B^{\phi_1+\phi_2}\;\;,
\end{equation}
where we have set,  $a_n={\rm const.}$. 

\begin{figure}[h]
\vspace{1pt} 
\includegraphics[width=0.5\textwidth]{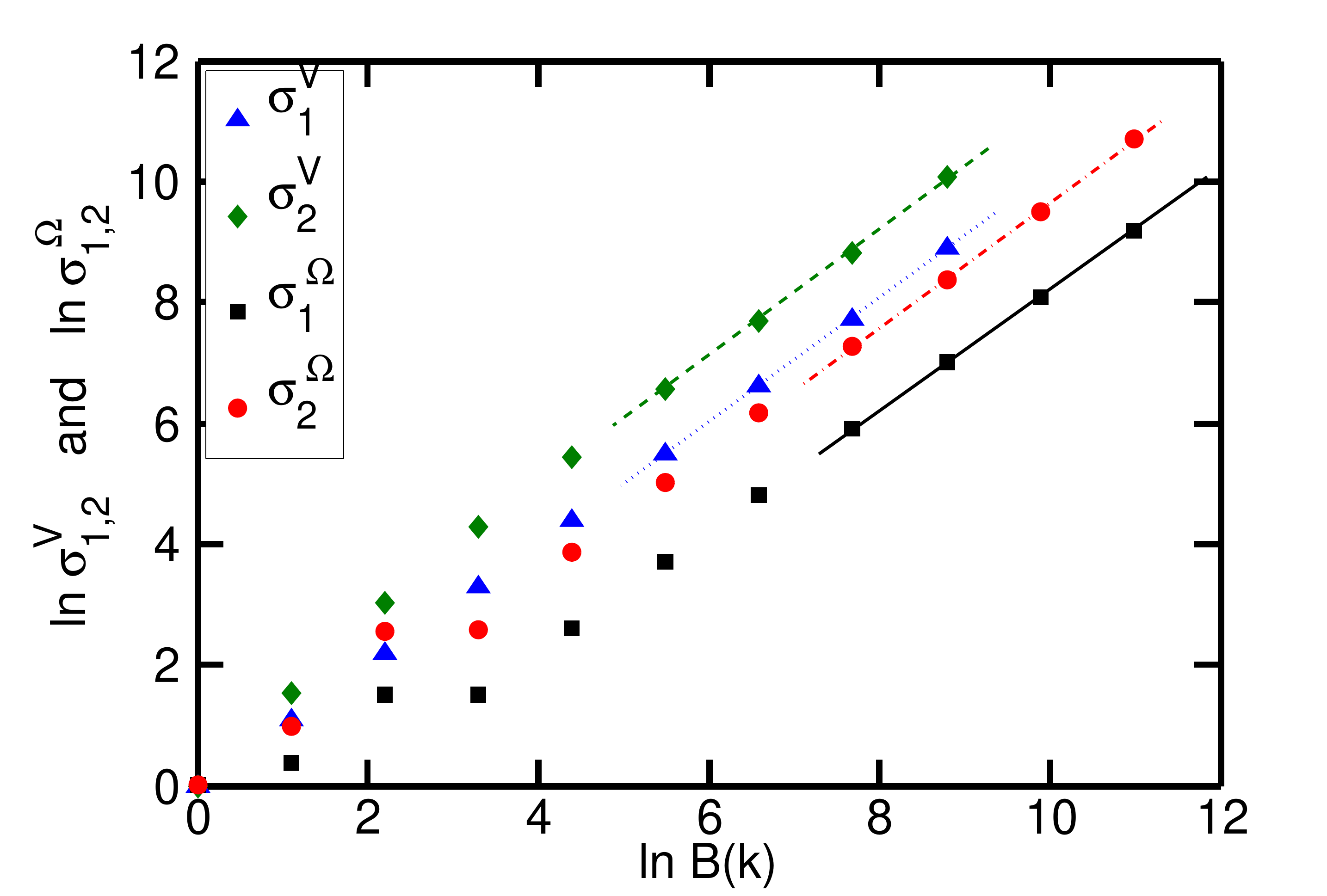}
\caption{(Color online) Rescaling factors $\sigma^{\rm V}_i$, and $\sigma^{ \Omega}_i$, $i=1,2$ for the Cayley tree with branching number $b=3$. See text, 
Eqs.(\ref{eq:sigma1},\ref{eq:sigma2},\ref{eq:sigma1O},\ref{eq:sigma2O}).
The scale factor is chosen as $B(k)=b^{k-1}$.
The linear fits  are to the last four points for each set. The exponents are found to be $\phi_1=1.00$, $\phi_2=1.03 \pm0.04$,  $p_1=1,\;\; p_2=1.01 \pm 0.01$.  The critical exponents are given in Table~\ref{allexponents}.  }
\label{sigma12Cayley}
\end{figure}

Doing the sums  for $r\gg 1$, $1\ll k < r$, (i.e. in the small $\om$ region), we find $\phi_1=\phi_2=1$. 
The numerically obtained scaling behavior of $\sigma^{\rm V}_1$ and  $\sigma_2^{\rm V}$, as well as $\sigma^{\Omega}_1$ and $\sigma_2^{\Omega}$  are shown in Fig.~\ref{sigma12Cayley}, and are  in agreement with our approximate analytical result.   The critical exponents are given in Table~\ref{allexponents}. 

\subsection{Gaussian model on the diamond lattice}
We next consider a  hierarchical lattice, sometimes also known as the diamond lattice~\cite{diamond}. The zeroth generation consists of two nodes connected by an edge;  at the first iteration the edge is replaced by a  rhombus with the two new nodes making up the first generation, and the network is constructed by iteratively replacing each edge of a rhombus by yet another rhombus. Indexing the different generations by $k=0, \ldots, r$, the total number of nodes is   $N_r=2(1+\sum_{k=0}^{r-1}4^k)$, the number added at each generation is  $N_k - N_{k-1} = 2^{2k-1} $ for any $k\ge 1$.  

After $r$ iterations, the nodes belonging the $k$th  generation have  degrees $d_k$, where  $d_0=d_1= 2^r$ and $d_k = 2^{r-k+1},$ for $ 1<k\le r$.  This leads to an overall scale free degree distribution  with $\gamma=2$. On the other hand, embedding the lattice in a metric space  and regarding the iterations as a fine-graining operation~\cite{GriffithsandKaufman} leads to a   multifractal degree   distribution over the lattice~\cite{Erzan1997}.  
\begin{figure}[ht]
\vspace{1pt} 
\includegraphics[width=0.5\textwidth]{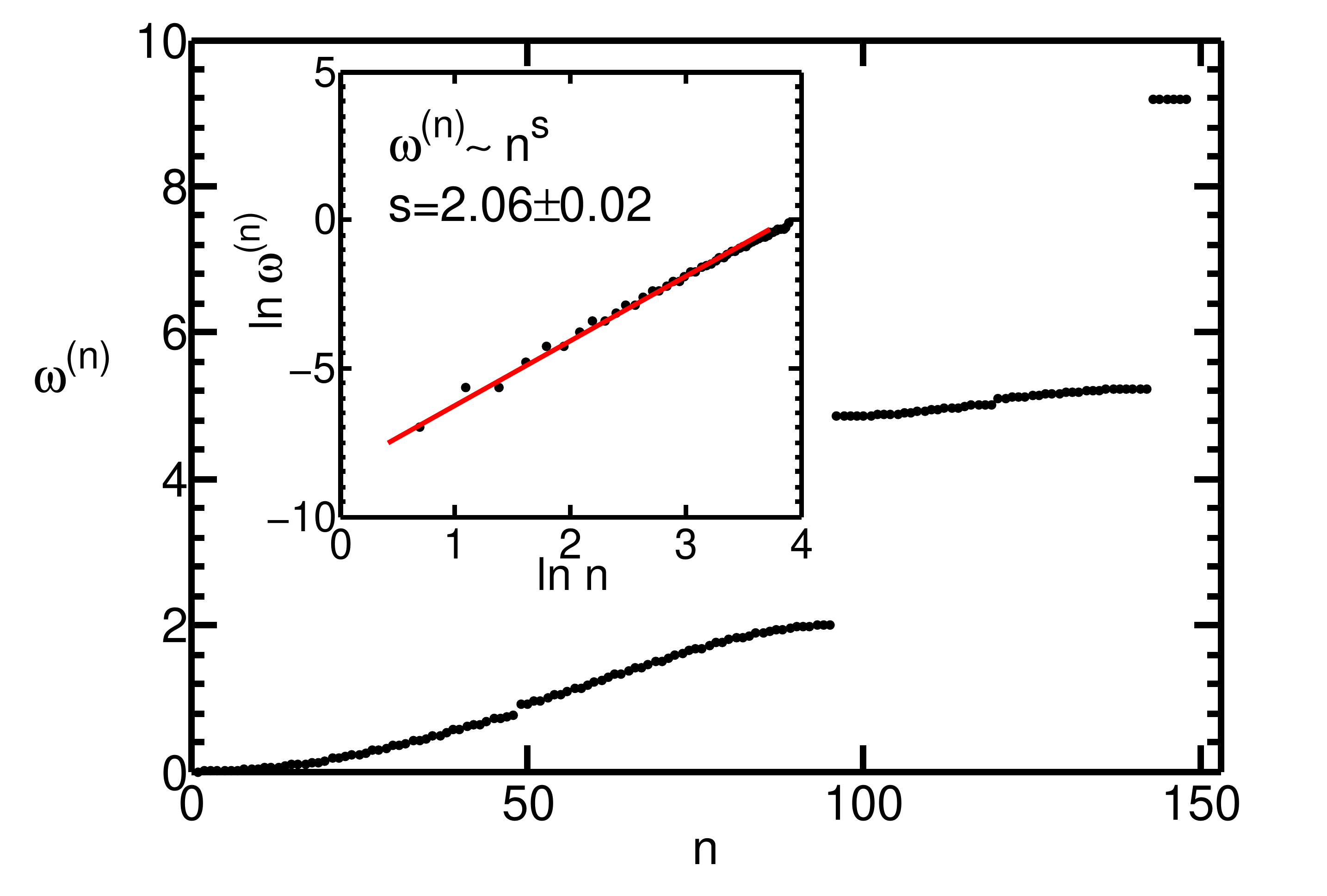}
\caption{(Color online) The distinct eigenvalues $\om^{(n)}$ of the graph Laplacian on the diamond lattice plotted against $n$.  The inset shows the scaling behavior in the small $\om^{(n)}$ region.}\label{fig:wnvsninset} 
\end{figure}
\begin{figure}[ht]
\vspace{1pt} 
\includegraphics[width=0.5\textwidth]{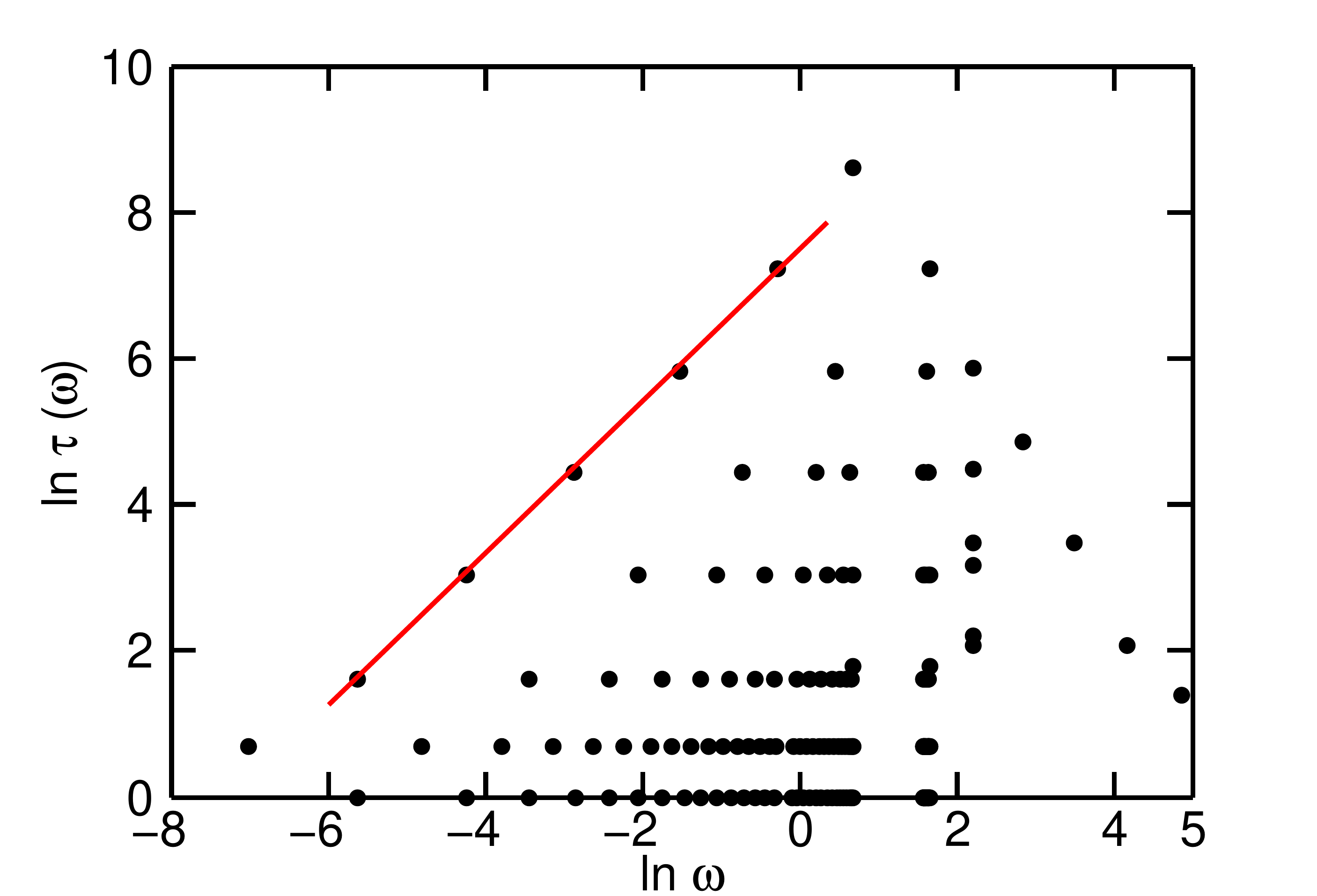}
\caption{Spectral distribution for the diamond lattice, for $r=7$ generations, on a log-log scale.  Families of eigenvalues with degeneracies $\tau_j=\sum_k^j 4^k$ can be seen as horizontal sets of points.}\label{fig:diamondrho} 
\end{figure}
\begin{figure}[ht]
\vspace{1pt} 
\includegraphics[width=0.5\textwidth]{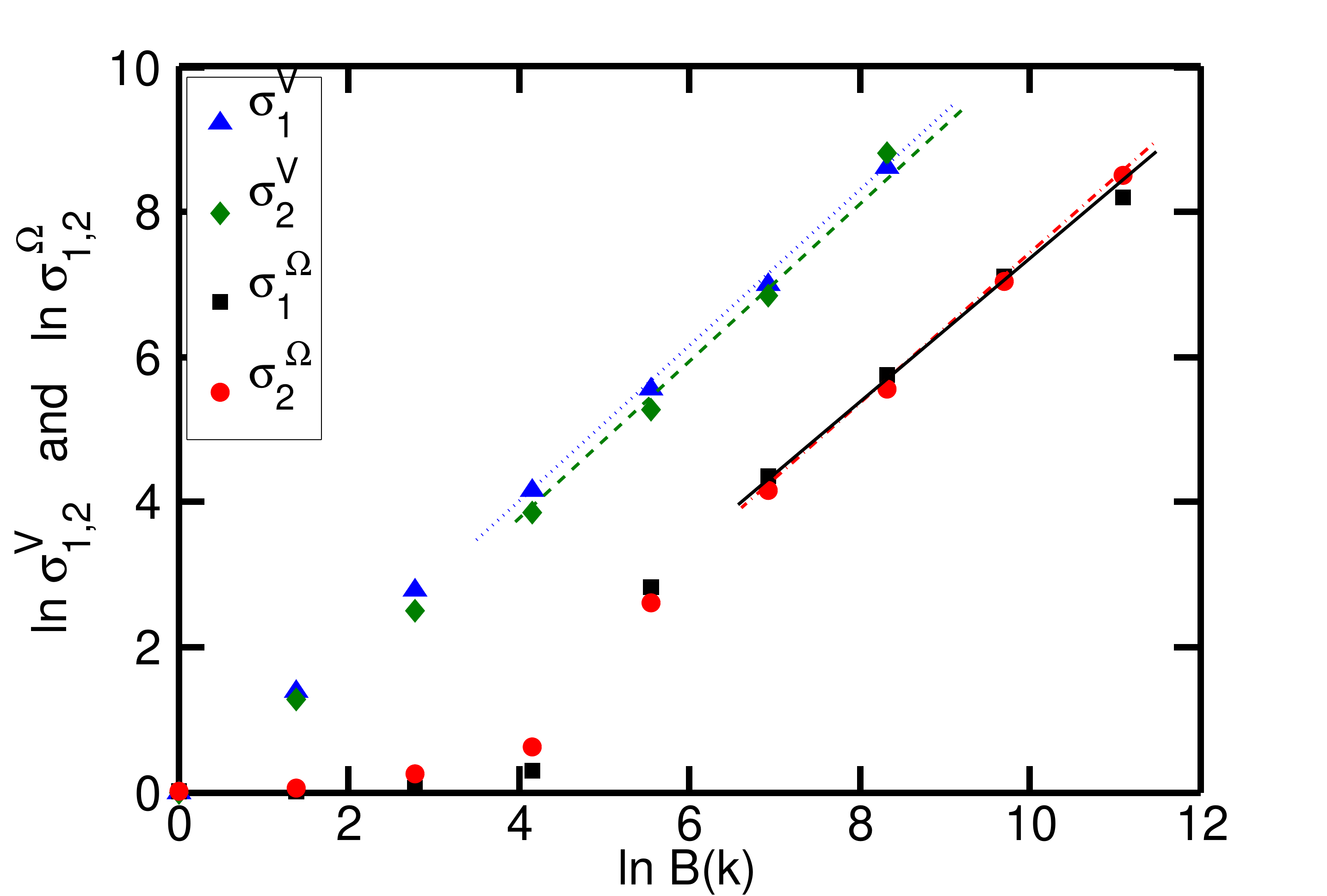}
\caption{(Color online) The numerically calculated rescaling factors for the diamond lattice. The scale  factors have been chosen as $B(k)=2^{2k}$ and $1\le k<\le r+1$ for $ \sigma_1^{\rm V},\;\; \sigma_2^{\rm V}$ and $1\le k \le r+2$ for  $\sigma_1^{\Omega},\;\; \sigma_2^{\Omega}$,  for a total number of generations $r=7$. The linear fits are to the last four points for each set. We get $\phi_1=1$,  $ \phi_2=1.01\pm0.03$, 
$p_1=0.97\pm 0.04$ and $ p_2=1.06\pm0.02$. 
The critical exponents are summarized in Table~\ref{allexponents}.}\label{fig:diamondfii} 
\end{figure}
We are interested in the region  $ 0\le \om \le 2$ where $\langle \om^{(n+1)}-\om^{(n)}\rangle\equiv \ell \to 0 $ as  $4^{-r}$.   We find that the degeneracies of the distinct eigenvalues  $\om^{(n)}$ obey a scaling relation $\om^{(n)}\sim n^s$ with $s={2.06}\pm 0.02$ for small $n$ (Fig.~\ref{fig:wnvsninset}).  The degeneracies $\tau_n$ are organized in triplets, $\bT_j =\{ 1,2,\tau(j)\}$, with $\tau(j)$ being defined now as the $j$th distinct element of the series $1,5,21,$ etc. given by $\tau(j+1)= 4\tau(j)+1$, or  $\tau(j)=\sum_{k=0}^{j} 4^k\sim 4^j$, for $0\le j \le r-1$.  The spectral distribution is naturally embedded in the real numbers and has a multifractal structure, which is generated by the replacements   $\bT_1 \to \{\bT_1 ,\bT_2, \bT_1\}$ and, for $  j\ge 2$,  $\bT_j \to \bT_{j+1}$  at each iteration. 

The spectral distribution for the diamond lattice, plotted on a log-log scale,  is shown in Fig.~\ref{fig:diamondrho}, with the initial $\om$ values for each $j$th family being given by $\om_{\rm init}(j) \propto 4^j$ obeying the scaling relation $\tau(j) \propto \om_{\rm init}(j)$, so that 
the envelope of the distribution is qualitatively the same as that  of the Cayley tree, Fig.~\ref{fig:Cayley_rho}.

The complexity of this spectral distribution is tamed by the  rescaling factors $\sigma_1$ and $\sigma_2$. It should be noted that  $\sigma_1^{-1} \sim \sum_n^{n_B} \tau_n $ and $\sigma_1^{-1}\sigma_2^{-1}\sim  \sum_n^{n_B} \om_{(n)}\tau_n $  (with $n_B=r-1-k$ for $B(k)=4^k$) are, respectively the (truncated) zeroth and 1st moments of the multifractal spectral distribution described above. These integrated quantities smooth out the singular nature of the distribution itself, and yield simple scaling law for large  $B$ (small $\om$), the relevant regime for the critical behavior.

In Fig.(\ref{fig:diamondfii}) $\sigma^{\rm V}_1$,  $\sigma^{\rm V}_2$, $\sigma^{\Omega}_1$ and $\sigma^{\Omega}_2$, are calculated numerically, again with a set of scale factors  $B(k)=4^{k}$, in keeping with the discrete scaling symmetry of the lattice. The exponents are given in  in Table~\ref{allexponents}. The numerically  computed values of the critical exponents  are in agreement with  the scaling behavior ($\beta=0$, i.e., ${\tilde d}=2$) which one may read off from the envelope of the spectral distribution (Fig.\ref{fig:diamondrho}).  
 \begin{table*}[!ht]
\caption{Exponents of the spectral density, the rescaling factors and critical exponents obtained 
for the Gaussian model on spatial and non-spatial networks.  The first set of exponents are obtained by the method of scaling the total number of modes, and the second set via scaling the upper cutoff for the eigenvalues.  The  values  for $\delta$ that are  larger than the inverse of the error bars around zero have been shown as $\infty$.  The Gaussian model yields identical results on the square lattice, the Cayley tree and the diamond lattice, which all have spectral dimension $\sd=2$.  The exact Gaussian values are indicated with the subscript $G$. See text for definitions.}
\begin{center}
	\begin{tabular}[c]{l c c c c c c c c c c c c}
		\hline\hline
Network           & 	$\beta$ &\quad $\phi_1$\quad&     \quad  $\phi_2$   \quad  &   \quad   $\alpha$ \quad  \quad    & \quad $\delta$ \quad    & \quad \quad  $p_1$  \quad \quad\quad        &     \quad \quad     $p_2$ \quad\quad     &  \quad \quad \quad$\alpha$ \quad   \quad &\quad  $\delta$ \quad&\quad  $\alpha_G$  \quad   &  \quad $\delta_G$   \\
		\hline
Square 	  &  $0.00\pm 0.02$   &     $1$   &$0.996\pm0.003$&          $1$          & $\infty$       & $1.00\pm0.01$ &  $0.99\pm0.03$ & $ 1.00\pm0.03 $  &  $ \infty$ &1&$\infty$\\
Cubic   	  &   $0.5\pm 0.1$	&     $1$   &  $0.66\pm0.05$ & $0.48\pm0.12$ & $4.88\pm0.25$   & $1.49\pm0.02$ &  $0.98\pm0.05$ & $ 0.48\pm0.05 $  &  $ 4.8\pm0.4$& $1/2$& 5\\
Cayley$_3$ &  $1.03\pm 0.04$	&     $1$   &  $1.03\pm0.04$ & $1.03\pm0.04$ & $\infty$   &          $1$          &  $1.01\pm0.01$ & $ 0.99\pm0.01 $  &  $ \infty$ &1&$\infty$\\
Cayley$_5$ &  $1.06\pm 0.09$ 	&     $1$   &  $1.06\pm0.09$ & $1.06\pm0.09$ & $\infty$   &          $1$          &  $1.05\pm0.05$ & $ 0.95\pm0.05 $  &  $ \infty$ &1&$\infty$\\
Diamond  	  &  $1.01\pm 0.03$ 	&     $1$   &  $1.01\pm0.03$ & $1.01\pm0.03$ & $\infty$ & $0.97\pm0.04$ &  $1.06\pm0.07$ & $ 1.09\pm0.11 $  &  $ \infty$&1&$\infty$\\
			\hline\hline
		\end{tabular}
	\end{center}
\label{allexponents}
 \end{table*}

\subsection{Recovering the Mean Field exponents}
Goldenfeld~\cite{Goldenfeld} discusses the anomaly of obtaining non-classical (non-mean field) values for the critical exponents of the Gaussian model, which is  based on a Landau expansion (see Eq.(\ref{eq:landau})) and points out that the anomaly can be understood in terms of the dangerous irrelevant field $v_0$. We may repeat the argument in the present case, using the scaling relation in Eq.~(\ref{eq:kadanoff}) with a third scaling field, $v$, so that $f=B^{-x} f^\prime (B^{Y_t(x)}t, B^{Y_h(x)}h, B^{Y_v(x)}v)$, yielding 
\be
m(0,h,v)=h^{[-Y_h(x)+x]/Y_h(x)} {\mathcal M}(v h^{-Y_v/Y_h}) \;\;, \label{eq:NG1}
\ee
for the magnetization on the critical isotherm. Here $x=1$ or $x=p_1=1+\beta$ depending on whether we scale the number of modes (Section~II.A) or the maximum eigenvalue (Section II.B), respectively. We will require that  $Y_v \propto 1-\beta$, in analogy with setting $y_v\propto 4-d$ in Euclidean space.  The Landau expansion gives $m(0,h,v)\propto (h/v)^{1/3}$; therefore one takes~\cite{Goldenfeld}  the scaling function ${\mathcal M}(v) \sim v^{-1/3}$ in the limit of small $h$.  Eq.(\ref{eq:NG1}) then gives 
\be 
-1 + \frac{x}{ Y_h(x)} +\frac{1}{3} \frac{Y_v(x)}{ Y_h(x)} = \frac{1}{\delta} \;\;.
\ee
For $x=1+\beta$, one has $Y_h=(2+\beta)/2$, $Y_v = 1-\beta$. 
For $x=1$, $Y_h=(2+\beta)/[2(1+ \beta)]$ and one must take  $Y_v=(1-\beta)/(1+\beta)$.  In either case,  $\beta$ cancels out of the final result,  yielding the mean field value for $\delta$.  
One may similarly show that keeping $v_0$ in the calculation and using the Landau expansion for $h=0$ to get $m\propto \sqrt{-r_0/v_0}$ gives  the order parameter exponent $\beta_m$ to be $1/2$, from which one may derive $\alpha=0$ using the scaling relation $\beta_m(1+\delta) = 2-\alpha$.  

\subsection{Square and cubic lattices}
For completeness, we have also computed the spectral densities of the square and cubic lattices, and their rescaling factors. 
The Laplace eigenvalues for the square and cubic lattices are analytically given by
\be
\om_{\bq} =  4 \sum_{j=1}^d \sin^2\left(\frac{q_j }{ 2}\right)
\ee
where we have indexed the eigenvalues by the wave vector, the lattice spacing is unity, $d$ is the Euclidean dimension and $q_j= (\pi n_j\, N^{-1/d})$ are the Cartesian components of $\bq$. In the limit of small $q=\|{\bq}\|$, $\om_{\bq} \simeq q^2$.  Then the spectral density is $\rho(\om) \propto \om^{\beta}$ with  $\beta= {d/ 2} -1 $.
 
The numerical results for the non-trivial scaling exponents and for $\alpha$ and $\delta$ are given in Table~\ref{allexponents}.
The plots of the spectral density and rescaling factors are provided in the Appendix.  It is instructive to compare the accuracy obtainable from the rescaling factors as opposed to the spectral densities themselves, which converge very slowly, in the small $\om$ region, to their thermodynamic limits. 

\section{Comparison with conventional methods}

In this section we would like to check our SRG results against conventional methods which we here adapt to non-spatial lattices, namely, exact summation of the leading term in the specific heat, obtained by differentialting Eq.~(\ref{eq:FEbeta}),  and finite size scaling (FSS) by the number of nodes of the lattice, instead of the linear size of the system.
\subsection{\label{sec:Exact} Exact Enumeration}
\begin{figure}[ht]
\centering 
\includegraphics[width=0.5\textwidth]{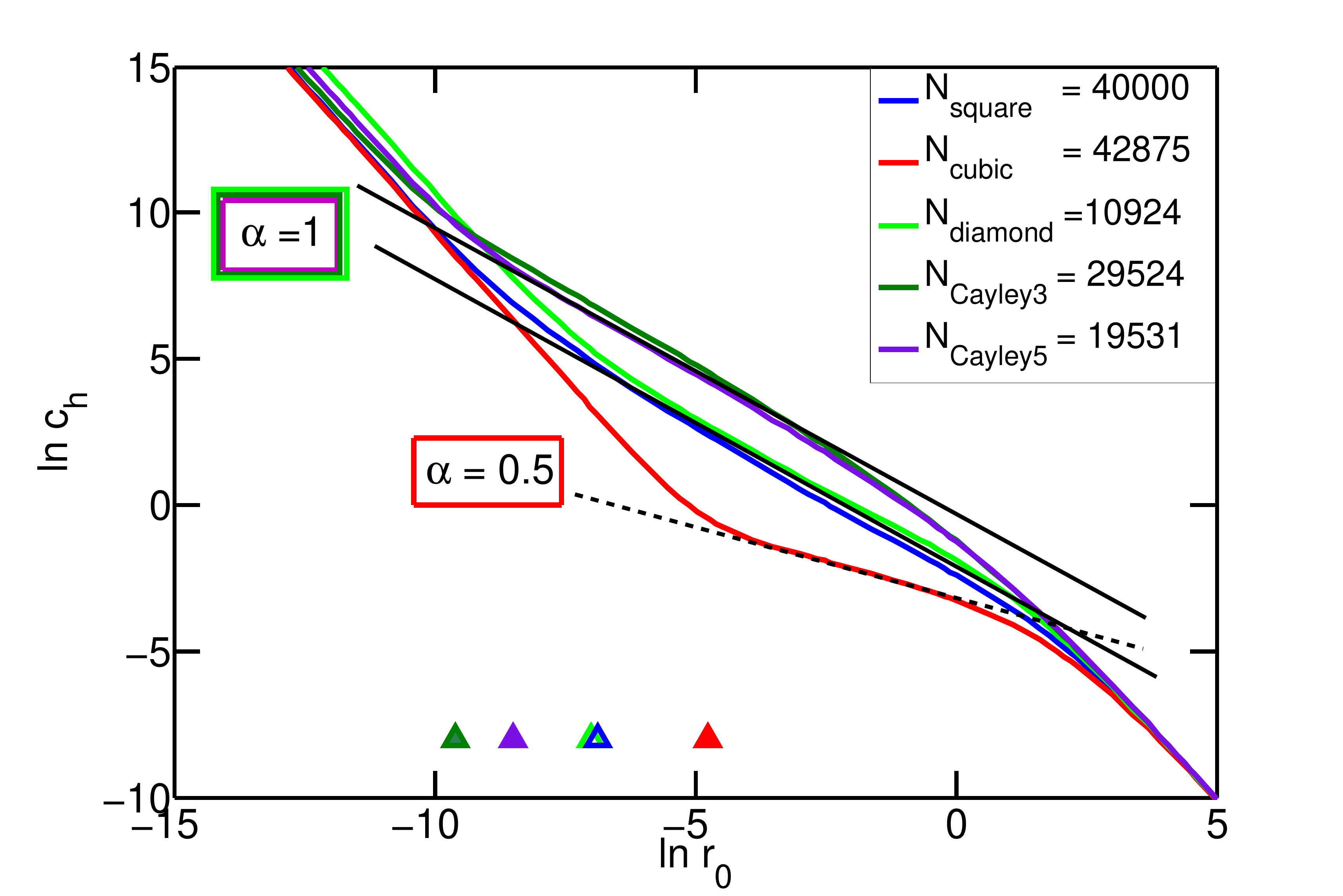}
\caption{(Color online) Specific heat exponent $\alpha$ obtained from Eq.(\ref{eq:cv}), for the Cayley tree with branching numbers $b=3,\;5$ (indistinguishable in graph) and the diamond, square and cubic lattices. The legend indicates the size of the different lattices.  The positions of the first nonzero eigenvalues of the graph Laplacian are indicated by triangles with colors that match the curves, and mark the onset of the scaling region. The slope of the tangent lines yield $\alpha=1$ for all except the cubic lattice, which has $\alpha=0.5$.}
\label{fig:C_exact}
\end{figure}

For the Gaussian model, the specific heat can be explicitly calculated to leading order as, 
\be
\label{eq:cv}
c_h \propto \frac{1}{2N} \sum_{\mu=1}^N \frac{1}{(r_0+\omega_\mu)^2}\\=\frac{1}{ N} \sum_{n=0}^{r-1} {\tau_n \over [t+\om^{(n)}]^2}\;\;.
\ee
The results for the Cayley tree, the diamond lattice, and square and cubic lattices, are shown in Fig.(\ref{fig:C_exact}).
The critical scaling behaviour of $c_h$ is obtained  for $r_0$ between the first nonzero Laplace eigenvalue and the van Hove singularity in the Laplacian spectral density (which falls near unity) and yields the critical exponent $\alpha$ in agreement with  the SRG results in Table~\ref{allexponents}.

\subsection{\label{sec:Fss}  Finite Size Scaling for non-spatial lattices}

In order to estimate possible errors due to the finiteness of the lattices considered, an FSS analysis adapted to non-spatial lattices was performed for the example of the Cayley tree. The size-dependent relevant effective ``field" in this case is    $N^{-1}$, in place of the linear 
size of the spatial lattice~\cite{Goldenfeld}.  

The specific heat should then scale as 
\begin{equation}\label{eq:Cvlimitson}
\frac{c_h(t,N^{-1})}{N^{2Y-1}} = \begin{cases} {\rm const.}, &\mbox 
			 {$x< 1$} \\
			x^{1-2Y}, & \mbox{$x > 1 $}.
			\end{cases}
\end{equation} 
where $x=t^{(1/Y)} N$.  Plotting $N^{-(2Y-1)} c_h$ vs. $Nt$, we find $Y=1$, with a satisfactory collapse for $r=3,\ldots,7$.
We can show that $\alpha=2-Y^{-1}$ and our SRG result for $\alpha$ is confirmed.


For the Cayley tree, the same result may be obtained by substituting  the approximate analytic results 
 for $\tau_n$ and $\om^{(n)}$ from Eqs.(\ref{omega},\ref{eq:tau-omega}) into  Eq.~(\ref{eq:cv}). Noticing that $b^r\sim N_r$, one gets, after multiplying and dividing the RHS by $N_r^2$ and simplifying,
\be \label{eq:cvtau1}
\frac{c_h}{N_r} \simeq \frac{b-1}{b^2}\sum_{n=0}^{r-1} \frac{b^n}{(N_r\,t+b^n)^2}\;\;.
\ee
The RHS is only a function of $N_r\,t$ and approaches a constant for $N_rt < 1$. Going over to an integral immediately gives the result that $c_h\sim (N_r\, t)^{-1}$ for $N_r\,t >1$.

\section{Including the  $\psi^4$ interactions}

In this section we will add a $\psi^4$  term (Eq.\ref{eq:landau}) to the Gaussian Hamiltonian and treat this system on the Cayley tree and the diamond lattice.  This interaction term  leads to couplings between different fluctuation modes, and the precise nature of the eigenvectors come to play an important role.  


The Ising model exhibits Mean Field critical behavior on the Bethe Lattice (Cayley tree in the infinite limit)~\cite{Eggarter,CJThompson}, therefore we should expect to find that on this network,  the  Gaussian fixed point is stable with respect to the inclusion of a $\psi^4$ coupling.  However,  on the diamond lattice, we would expect the emergence of a non-Gaussian fixed point as well, since the Ising model on the diamond lattice undergoes an order-disorder phase transition with non-trivial exponents.~\cite{KaufmanandGriffiths,GriffithsandKaufman}

 
 The interaction term  $H_{\rm int}=v_0\sum_i \psi^4(i)$ in the Hamiltonian (see Eq.~\ref{eq:landau}) can be expanded explicitly in terms of the eigenvectors of the graph Laplacian  on an arbitrary network, as
\be \label{eq:psi4_1}
H_{\rm int} = v_0\sum_{1,2,3,4} \hpsi_{1}\hpsi_{2 }\hpsi_{3}\hpsi_{4} \;\Phi(1,2,3,4) \;\;,
\ee
where, for brevity,  we have written $\{1,2,3,4 \}$ instead of  $\{ \mu_1,\mu_2,\mu_3,\mu_4\}$  and we have defined the 4-vertex
\be
\label{eq:Phi} 
 \Phi(1,2,3,4) = \sum_i^N u_{1}(i) u_{2}(i)u_{3}(i)u_{4}(i) \;\;\;.
\ee
Here $u_{\mu}(i) $ is the $i$th element of the eigenvector $\mathbf{u}_\mu$ of the graph Laplacian. 

On a periodic lattice where the eigenvectors are the harmonic functions  $u_{\mu}(j)=N^{-1/2} \exp(i\bk_\mu \cdot \bx_j) $ and $\omega_\mu = \| \bk_\mu \|^2 $, one immediately has 
\be
\Phi(1, 2,3,4) = \delta^{(d)} (\bk_1+\bk_2+\bk_3+\bk_4)\;\;,
\ee
where, in the thermodynamic limit  $\delta^{(d)}$ becomes the $d$-dimensional Dirac $\delta$-function. In the case of an arbitrary network, a similar constraint is difficult to find in general, even with an analytical solution for the eigenvalues~\cite{erzan,Rojo,Rabbiano,Burda}.  In this paper we will avail ourselves of the numerically calculated eigenvectors for the respective lattices.

We now explicitly perform the scaling with respect to the maximum eigenvalue, by taking  partial integrals in the partition function, over fields $\hpsi_\mu$ with $\mu > \mu_B$,   with the Gaussian weight $e^{-H_0^>}$. We choose to define $\mu_B$ as in  Eq.(\ref{muBOmega}), so that we are explicitly  truncating the largest eigenvalue; however, truncating the number of modes gives parallel results for $\sd = 2$.  
The superscripts $>$ and $<$ have the same meaning as in Section II,  i.e., $H_0^> =  \frac{1}{2}\sum_{\mu > \mu_B}^{N}[r_0+\omega_\mu]\hpsi_\mu^2$.
Note that the functional $ e^{-H_{\rm int}}$ involves fields $\hpsi_\mu$ with $\mu$ in both the lower and upper ranges with respect to the cutoff $\mu_B$,
 We obtain, 
\be \label{eq:partial}
Z(r_0, v_0)= Z_0^> \int_{-\infty}^\infty \prod_{\mu<\mu_B} d\hpsi_\mu e^{-H_0^<} \langle e^{-H_{\rm int}} \rangle_0^>\;\;.
\ee
The normalization factor  is  $Z_0^>=  \int_{-\infty}^\infty \prod_{\mu>\mu_B} d\hpsi_\mu e^{-H_0^>}$,   and we have implicitly defined the expectation value,
\be \label{eq:upperexp}
\langle {\cal Q}  \rangle_0^> = (Z_0^>)^{-1}  \int_{-\infty}^\infty \prod_{\mu_B<\mu<N} d\hpsi_\mu {\cal Q} e^{-H_0^>}\;\;\;.
\ee
The perturbation up to second order in the coupling constant is obtained via a cumulant expansion,
 \be \label{eq:cumulant}
\langle e^{-x} \rangle \simeq e^{-\langle x \rangle} e^{\frac{1}{2}[\langle x^2\rangle - \langle x\rangle ^2]}\;\;.
\ee
\subsection{First order terms}
\begin{figure}[ht]
\vspace{1pt} 
\centering 
\includegraphics[width=0.4\textwidth]{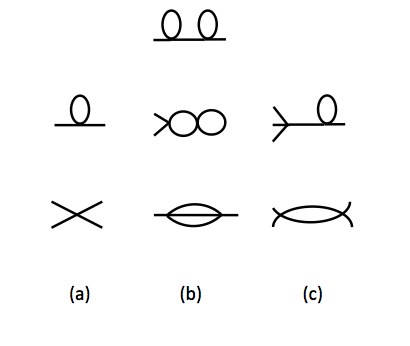}
\caption{The Feynman diagrams for the quadratic and quartic interactions, up to second order in the coupling constant. The labels used in the text are column (a), $D_{2,1}$, $D_{4,1}$, in column (b) $D_{2,2}$, $D_{2,2^\prime}$  $D_{2,2^{\prime\prime}}$,  in column (c), $D_{4,2^\prime}$ and $D_{4,2}$.} \label{fig:Feynman}
\end{figure}

The different diagrams corresponding to the different terms in the perturbation expansion are given in Fig.~\ref{fig:Feynman}.
There are only two  terms arising from the cumulant expansion  to first order in $v_0$, and they are,
\be
\label{eq:H21}
H_{2,1}^< = 6  v_0 \sum_{1,2<\mu_B} \hpsi_{1} \hpsi_{2} 
                 \sum_{3 > \mu_B} G(\om_{3},r_0) \Phi( 1, 2, 3, 3)  
\ee
and
\be
H_{4,1}^< = v_0 \sum_{1, 2, 3, 4 <\mu_B}  \hpsi_{1}\hpsi_{2 }\hpsi_{3}\hpsi_{4} \Phi(1, 2, 3, 2)                       
\;\;.   
\label{eq:H41}
\ee

The Green's function $G(\om, r_0)$ arises from the contraction of two fields with indices  $\mu> \mu_B$ and  is defined via 
\be
\langle \hpsi_\mu \hpsi_{\mu^\prime}\rangle_0^> = \delta_{\mu,\mu^\prime} \frac{1}{\om_\mu + r_0} \equiv \delta_{\mu,\mu^\prime} G(\om_{\mu}, r_0) \;\;.
\label{eq:correl}
\ee

We see from Eqs.~(\ref{eq:H21},\ref{eq:H41}) that the couplings have acquired an eigenvector dependence, in a way similar to the case on periodic lattices.  For the Cayley tree we can explicitly show that the 4-vertex   $\Phi( \mu_1, \mu_2, \mu_3,\mu_3)$ is nonzero  only if its arguments are {\it a)} either equal pairwise, or {\it b)} one of them is equal to the constant vector and the remaining three equal to each other.  In obvious notation, 
\begin{multline*}
\sum_{ \{ 1,2,3,4\}} \Phi ( \mu_1, \mu_2, \mu_3,\mu_4)\\
\; ={{4}\choose{2} }\sum_{ \{ 1,2,3,4\}} \delta_{\mu_1,\mu_2}\delta_{\mu_3,\mu_4} \Phi(\mu_1, \mu_2, \mu_3,\mu_4) +\\
 \;+ 4 \sum_{\{ 1; 2,3,4\}}\delta_{\mu_1,1}\delta_{\mu_2,\mu_3,\mu_4}  \Phi( \mu_1, \mu_2, \mu_3,\mu_4) \;\;.
\end{multline*}
In Eq.(\ref{eq:H21}), the requirement that $1,2 \le \mu_B$ and $3,4 > \mu_B$ means that only the first set of conditions {\it  (a)} can be satisfied and therefore $H_{2,1}^<$   is actually diagonal in $\mu_1, \mu_2$  and contributes to the quadratic term in the truncated Hamiltonian.  
For the diamond lattice we find the same result, numerically. 

It is  convenient to define 
\be 
I_1(B, r_0, i) = \sum_{\mu= \mu_B+1}^N G(\om_\mu,r_0) u_\mu^2(i) \;\;,
\label{eq:I1i}
\ee
and
\be 
I_1(B, r_0) = \sum_i^N I_1(B, r_0, i) 
\label{eq:I1}
\ee

In addition to the scaling factors $\sigma_1^\Omega, \; \sigma_2^\Omega$, we now have to also define   $\sigma_{2,1}$  and  $\sigma_{4,1}$ for the first order   quadratic and quartic  terms (where we have dropped the superscript $\Omega$), 
\be 
\label{eq:sigma21}
\sigma_{2,1}(B) \equiv  \frac{\sum_{\mu_1,\mu_2}^{N}   \sum_{i=1}^N I_1(1,i) u_{\mu_1} (i) u_{\mu_2} (i) }{\sum_{\mu_1,\mu_2}^{\mu_B}   \sum_{i=1}^N I_1(B,i) u_{\mu_1} (i) u_{\mu_2} (i) } \sim B^{\phi_{2,1}}\;\;,
\ee
and
\be 
\label{eq:sigma41}
\sigma_{4,1}(B) \equiv \frac{\sum_{1,2,3,4}^{N} \Phi(1,2,3,4)}{\sum_{1,2,3,4}^{\mu_B} \Phi(1,2,3,4)}\sim B^{\phi_{4,1}}\;\;.
\ee
These scaling factors have been computed at $r_0=0$. The numerical values of the scaling exponents for  the Cayley tree and the diamond lattice  are given in Table~\ref{t:Feynman}.


To first order in $v_0$, we get, after rescaling the effective Hamiltonian and neglecting the explicit eigenvector dependence in the second term,
\begin{eqnarray*}
H_{1}^\prime &=& \frac{1}{2} \sum_{\mu=1}^N [r_0 B^{-p_1}+B^{-p_1-p_2} \om_\mu]z^2\hpsi^2 \\
                           +&& 6v_0 \sum_{\mu=1}^N  z^2 \hpsi_\mu^2 
B^{-\phi_{2,1}} \\
                            +&& v_0 \sum_{1,2,3,4} z^4 \hpsi_1 \hpsi_2\hpsi_3\hpsi_4  \Phi(1,2,3,4) B^{-\phi_{4,1}} -h z\psi_1 \;\;,
\end{eqnarray*}
where we have suppressed all numerical coefficients, as we will do in the rest of this presentation. 

Expanding $I_1(B,r_0)$ in $r_0$,  in order to obtain the linear contribution to the recursion relation of $r_0$, we write $\sigma^{-1}_{2,1}\sim B^{-\phi_{2,1}}\sim B^{-\phi_{2,1}^\ell}[I_1^{(0)}- r_0\, I_2^{(0)}]$, where $\ell$ signifies a summation over the outer legs, while the terms in the brackets are given by the loop integral expanded in terms of $r_0$. We find $I_1^{(0)}(B)\sim \log B$ and $I_2^{(0)}(B)\sim B^{\phi_{I2}}$ (see Table~\ref{t:Feynman}). 
Since we computed $\sigma^{-1}_{2,1}$ at $r_0$, we compare it with $B^{-\phi_{2,1}^\ell}I_1(0)$ and see that $\phi_{2,1}$ is consistent with 1 minus a small number. What appears as a small exponent actually corresponds to a logarithmic factor.
The factor $z$ is found again from Eq.~(\ref{eq:zOmega}) and does not change in the first order calculations.
%
%

Using Eq.\ref{eq:zOmega} and the results of Section II.A,  the recursion relations for $r$ and for $v$ to first order are,
\be
r^\prime=B^{p_2} r_0 + 12 v_0 z^2 B^{-\phi_{2,1}^\ell}  [I_1^{(0)} - r_0 I_{2}^{(0)}] \;\;.
\label{eq:rprime}
\ee
The third term in Eq.\ref{eq:rprime} is of second order in the small quantities $r_0,\; v_0$ and may be dropped. If we  ignore the eigenvector dependence, we can define, 
\be
 I_K^{(0)} = \sum_{\mu=N/B}^N \frac{1} {\omega_\mu^K}\;\;.
\ee
and are then able to provide analytical estimates for the scaling behavior of these functions.  See the  Appendix for the computation in the case of the Cayley tree, where we find $I_1(0)\sim \ln B$ and $I_2(0)\sim B$. 

To first order the recursion relation for $v^\prime$ is, 
\be
v^\prime= v_0 z^4 B^{-\phi_{4,1}} = v_0 B^{4-\phi_{4,1}}\;\;,
\ee
yielding $v^\ast= 0, \; r^\ast=0$ as the only fixed point; i.e., the Gaussian fixed point is stable to this order for both the Cayley and the diamond lattices, as we would expect from our experience with Bravais lattices~\cite{Goldenfeld}.  

%
\begin{figure}[ht]
\vspace{1pt} 
\centering 
\includegraphics[width=0.5\textwidth]{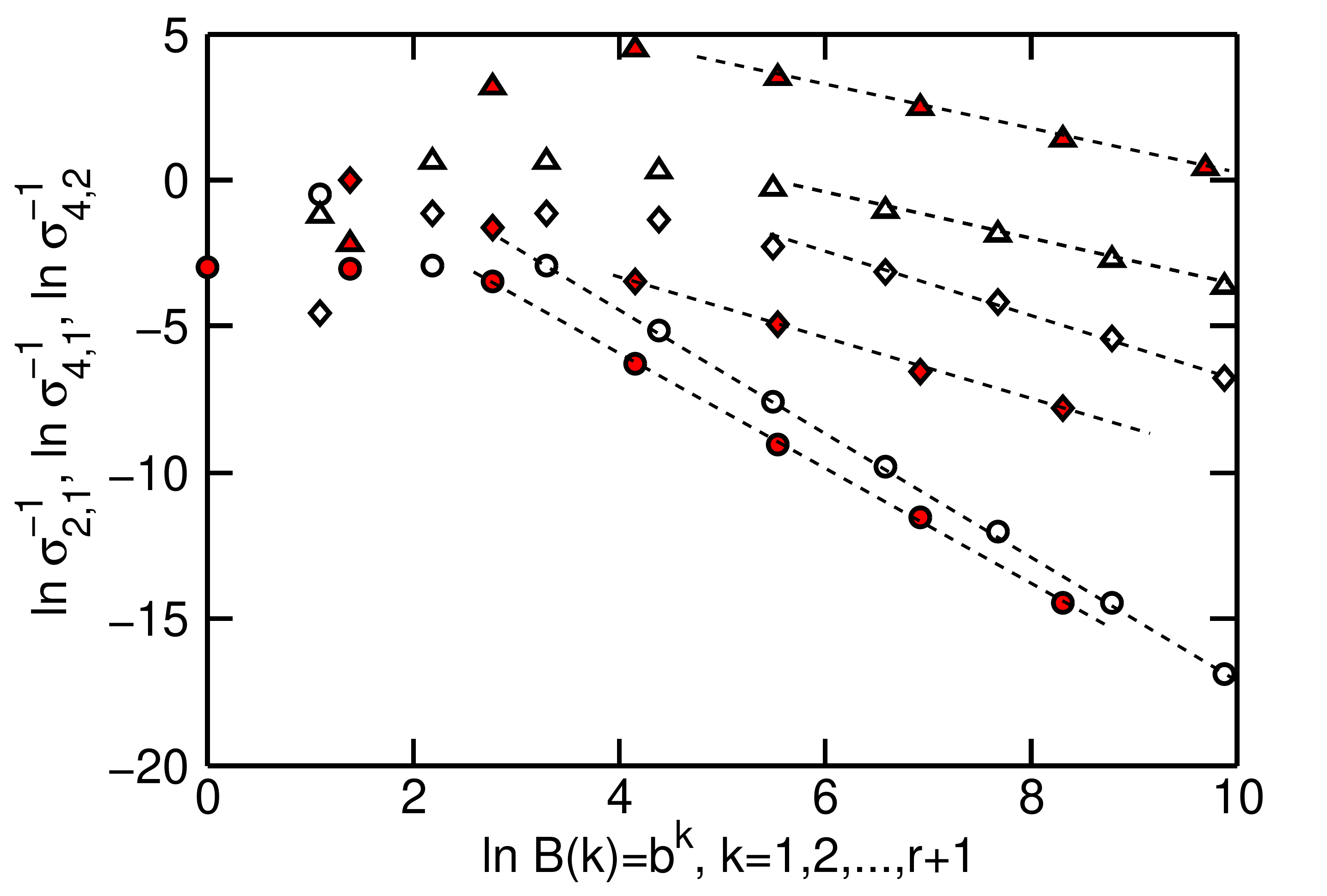}
\caption{(Color online) The renormalization group factors   $\sigma_{2,1}$,  $\sigma_{4,1}$, and  $\sigma_{4,2}$, for the Cayley tree (empty triangles, circles and diamonds, respectively) and the diamond lattice (red filled triangles, circles and diamonds),  in arbitrary units.  The points for the bubble diagram,  $\sigma_{4,2}$, for the Cayley tree have been shifted down by two decades for better visibility. The linear fits are indicated by dashed lines.} \label{fig:Cayley_diamond_scaling}
\end{figure}

\subsection{Second order terms}
To second order, there are more diagrams to consider.  
In the recursion relation for $r_0$  (Eq.~\ref{eq:rprime}), we have expanded in $r_0$, which is small by assumption, and we will keep only terms that are at most second order in the small quantities $v_0$ and $r_0$.  In the context of the $\epsilon$-expansion~\cite{Wilson3}, to first order in $\epsilon$ one could  ignore $v_0^2$  contributions to  Eq.(\ref{eq:rprime}), by arguing that they would be of higher order in $\epsilon$.  But here we still have to consider such terms that are  zeroth order in $r_0$.

The diagrams $D_{2,2\prime}$ and $ D_{2,2^{\prime\prime}}$ have vanishing amplitudes, nearly ten orders of magnitude smaller than the other terms, and go like $\log B$, so we neglect them. For the Cayley tree, as explained below   Eq.(\ref{eq:correl}), it is not possible to get    a nonzero contribution from  
$D_{2,2^{\prime\prime}}$ except for the case where one has the constant vector for the outer legs and three identical internal lines, giving only a negligible amplitude. 
The only diagram which could contribute to the second term in Eq.(\ref{eq:rprime}) is $D_{2,2}$ evaluated at $r_0=0$,  so we get,
\begin{multline}
r^\prime=B^{p_2} r_0 + 12 v_0 z^2 [B^{-\phi_{2,1}^\ell} 
   (I_1^{(0)} - r_0 I_{2}^{(0)}) \\ 
 - 6 v_0 B^{-\phi_{2,2}}]  \;\;.
\label{eq:rprime22}
\end{multline}
%

\begin{table}[!ht]
\caption{All the different terms which we consider to second order in the perturbative calculation of the renormalized  Hamiltonian on the Cayley tree and the diamond lattice.  The diagrams $D_{2,1}, \ldots D_{4,2}$, which are displayed in Fig.~\ref{fig:Feynman}, have been evaluated at $r_0=0$ and we scale the maximum eigenvalue $\Omega$. }  
\begin{center}
\begin{tabular}[c]{l c c c c }
\hline\hline
 Scaling Bhv. \quad  &\quad  Cayley \quad   & \quad Diamond \quad \\
\hline
$\sigma_{2,1}\sim B^{\phi_{2,1}}$ &  $\phi_{2,1}=0.80\pm0.03$ \quad& $\phi_{2,1}=0.75\pm0.02$ \\
$I_1^{(0)}(B) $ &  $\log B$ \quad& $\log B$ \\
$I_2^{(0)}(B)\sim B^{\phi_{I2}} $ & $ {\phi_{I2}} ={1.09\pm0.04}$ \quad& ${\phi_{I2}}={0.92\pm0.03}$ \\
$\sigma_{4,1}\sim B^{\phi_{4,1}}$ &  $\phi_{4,1}=2.11\pm0.03$ \quad& $\phi_{4,1}=1.97\pm0.04$   \\
$\sigma_{2,2}\sim B^{\phi_{2,2}}$ &  $\phi_{2,1}=0.40\pm0.04$ \quad& $\phi_{2,1}=0.60\pm0.03$ \\
$\sigma_{4,2}\sim B^{\phi_{4,2}}$ &  $\phi_{4,2}=1.11\pm0.03$ \quad& $\phi_{4,2}=1.11\pm0.04$  \\
			\hline\hline
		\end{tabular}
	\end{center}
\label{t:Feynman}
 \end{table}
%

The only nontrivial one-loop  contribution to the 4-vertex  at second order comes from the bubble diagram $D_{4,2}$, since $D_{4,2^\prime}$ turns out to have a vanishing amplitude as well.  Let us define, 
\begin{multline*}
I_2(B, i, j; r_0) \equiv \sum_{\mu,\mu^\prime= \mu_B+1}^{N} G(\mu) G(\mu^\prime)\\
\times  u_\mu(i)u_\mu(j)u_{\mu^\prime}(i)u_{\mu^\prime}(j)\;\;.
\label{eq:I2}
\end{multline*}
where the $r_0$ dependence of the Green's functions are implicit. The leading contribution from the bubble diagram to the $\psi^4$ term is, 
\begin{eqnarray*}
H_{4,2}^< &=&  36 v_0^2 \sum_{1,2,3,4}^{\mu_B} \hpsi_1\hpsi_2 \hpsi_3\hpsi_4 \\
                &&  \times  \sum_{i,j=1}^N I_2(B,i,j;0)  u_1(i)u_2(i)u_3(j)u_4(j)\;\;.
\end{eqnarray*}
where $r_0$ has been set to zero and we will drop it from the notation. To rescale this term we define 
\begin{multline}
\sigma_{4,2}^{-1} \propto
 \sum_{1,2,3,4}^{\mu_B}  \sum_{i,j=1}^N I_2^{(0)}(B,i,j)  u_1(i)u_2(i)u_3(j)u_4(j)\\
 \sim B^{-\phi_{4,2}} \;\;.
\end{multline}

Putting together all the quartic terms we  get the recursion relation for the interaction constant $v$ up  to second order,
\be
v^\prime = v_0 z^4 [ B^{-\phi_{4,1}}  -36 v_0\, B^{-\phi_{4,2}}]\;\;.
\ee 
Recalling that $z^2=B^{p_1+p_2}$ (Eq.~\ref{eq:zOmega}), and that in the present case, with $\beta=0$,  $p_1=p_2=1$, 
\be 
v^\prime = v_0 B^{4- \phi_{4,1}} [1  -36 v_0\, B^{-\phi_{4,2} +\phi_{4,1}}]\;\;,
\label{vprime}
\ee
where we have again suppressed all numerical coefficients.  

 If one neglects the explicit eigenvector dependence of  $\sigma_{4,2}^{-1}$ and performs the  sums over the lattice points, the bubble diagram essentially factorizes into the integrals over the external legs and the loop integral  $ I_2^{(0)}\sim B$, (Eq.\ref{eq:I2}), so that   $\sigma_{4,2}^{-1} \sim  B^{-\phi_{4,1}} I_2^{(0)} $.    The scaling behavior  of all the various diagrams appearing in Fig.~\ref{fig:Feynman} are shown in  Fig.~\ref{fig:Cayley_diamond_scaling}.  The values of the different exponents are given in Table~\ref{t:Feynman}.  From Table~\ref{t:Feynman} we see  that  indeed $\phi_{4,2}\simeq \phi_{4,1}-1$, bearing out our estimate for both the Cayley tree and the diamond lattice.

Using this estimate, finally we may write down the recursion relation 
\be
v^\prime = v_0  B^{4-\phi_{4,1}} (1 -  36 v_0  I_2^{(0)} ) \;\;. 
\ee
which yields the fixed point equation
\be 
v^\ast = \frac{1-B^{-(4-\phi_{4,1})}} {36  I_2^{(0)}}\;\;.
\ee
Note that the scaling exponents involved in Eq.~(\ref{vprime})  are, within error bars, identical for the Cayley and diamond lattices.  In both cases, large $B$ (the infrared regime)  leads to the Gaussian fixed point once again since $ I_2^{(0)}\sim B$.  Iterating the recursion relation Eq.~(\ref{vprime}) leads to the same result.  

Although the Cayley tree has a spectral dimension ${\tilde d}=2$, which is the lower critical dimension for models with Ising symmetry, it is well known that the Ising model on the Cayley tree has mean field behavior on the Bethe lattice (the limit of $r\to \infty$ limit of the Cayley tree)~\cite{Eggarter,CJThompson}.  In fact a tree structure is a lattice on which the Bethe-Peirls approximation is exact, in the same way as the RSRG~\cite{newKadanoff,Kadanoff,Leeuwen} is exact for the diamond lattice\cite{KaufmanandGriffiths,GriffithsandKaufman}. We should therefore expect that on the Cayley tree, the Guassian fixed point should be stable with respect to  perturbation by a $\psi^4$ term and that is indeed what we find.   

The Ising model on the diamond lattice,  on the other hand,  which has the same spectral dimension as the Cayley tree, exhibits nonzero magnetization below $T_c>0$ and non-classical exponents~\cite{Bleher}.  We would have expected that the perturbation by a $\psi^4$ term would lead to   non-classical behavior for the diamond lattice, but a non-Gaussian fixed point eludes us.  
For Eq.~(\ref{vprime}) to yield a nonzero fixed point, with at most a logarithmic correction, one should have $4-\phi_{4,1}$ equal to a small constant, which one may use as an expansion parameter, and $I_2^{(0)}$ to have at most a logarithmic dependence on $B$. Note that on a Bravais lattice in $d=2$ dimensions,  precisely the same scenario would have led to a null result within the usual Wilson momentum shell renormalization group as well. 

\subsection{Spectral dimension $\sd\, >2$}

\begin{figure}[!ht]
\vspace{1pt} 
\includegraphics[width=0.3\textwidth]{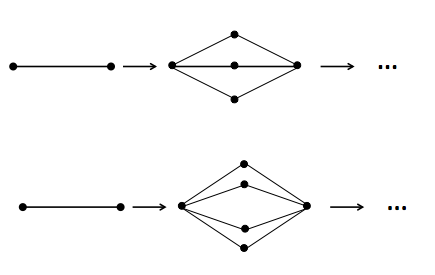}
\caption{The generalized hierarchical lattice a l\`a Itzykson and Luck~\cite{Itzykson}, shown for $p=3$ and $p= 4$ parallel paths with $b_s=2$ steps each.}\label{fig:pb_lattice} 
\end{figure}

In order to see whether we can obtain a non-Gaussian fixed point  for the quartic coupling constant $v_0$ and non-trivial critical exponents for $\sd\,>2$, we have utilized a 
generalization of the diamond lattice~\cite{Itzykson} illustrated in Fig.~\ref{fig:pb_lattice},  with  $p$ parallel paths replacing a bond and each  path consisting of $b_s$ steps.  This yields a fractal dimension of $d_f=\ln (p\,b_s)/\ln b_s$.  We take $b_s=2$ as in the foregoing sections, but vary $p$ and numerically calculate the spectral dimension $\sd$ from the scaling exponent $\beta$ in the region of small $\om$.  The relevant exponents  $\phi^{\Omega}_{4,1}$, $\phi^{\Omega}_{4,2}$, $\phi^{\Omega}_{4,2^\prime}$ and $\zeta^{\Omega} \equiv 4 \ln z/\ln B$ are provided in Table~\ref{t:bigger_d}.  In this section we scale the maximum eigenvalue and therefore we will again omit the superscript $\Omega$ in the remainder of the section, although we keep it in the tables for clarity.  
In dimensions $\sd>2$, we have not attempted to factorize the contributions from the outer legs and the internal loops, as we do not have a sufficient understanding of how the vertex $\Phi$ behaves.

Within our perturbative scheme up to second order in $v$, Eq.~(\ref{vprime}) is in the form of a  one-dimensional iterative map which can be written as
 $v^\prime = a_1 v (1-a_2 v)$, with   $a_1=B^{\zeta - \phi_{4,1}}$ and $a_2=36 \, B^{\phi_{4,1}}[B^{-\phi_{4,2}}+\frac{4}{3}B^{-\phi_{4,2^\prime}}]$, where, for $p>2$ we  include the diagram $D_{4,2^\prime}$ as it has a non-vanishing amplitude and an exponent close to that of  $D_{4,2^\prime}$.  The iteration converges to a  stable non-Gaussian  fixed point $v^\ast=(a_1-1)(a_1 a_2)^{-1}$ in the interval $v^\ast \in (0,a_2^{-1})$ provided $1<a_1<3$.  

We can read off from Table~\ref{t:bigger_d} that at $p=2$, the trajectory of $v$ is chaotic and goes off to infinity as the maximum of the $v^\prime$ curve exceeds $1/a_2$.  A non-Gaussian fixed point exists and is stable for $p=3,4,5$,  while for $p=7$,  $\sd = 4.12>4$,  we find    $a_1=3.5>3$, so the non-zero fixed point looses its stability. 

\begin{table}[!ht]
\caption{Higher spectral dimensions: Generalized diamond lattice (Fig.~\ref{fig:pb_lattice}). The spectral density exponent $\beta$, spectral dimension $\sd $, the exponents for the  field renormalization factor $z^4$ and for the Feynman diagrams contributing to  the  quartic coupling constant, as well as  the coefficient $a^\Omega $ and the stable fixed point  $v^\ast$ are provided for different $p\le7$;  for $p\ge 7$,  $\sd>4$. See text for definitions.}  
\begin{center}
\begin{tabular}[c]{c c c c c c c c c }
\hline\hline
\quad $ p$\quad\quad & \quad$\beta$\quad  &\quad  $\sd$ \quad & \quad $\zeta^\Omega$\quad  & \quad $ \phi^\Omega_{4,1}$ \quad & \quad $\phi^\Omega_{4,2}$ \quad &  $ \phi^\Omega_{4,2^\prime} $ & $a_1^\Omega $ \quad& \quad $v^\ast$ \quad  \\
\hline
2 &   0        &     2           &      4       &        1.99   &      1.11       & 1.425    &4   &0      \\
3 &    0.40   &     2.80     &     4.80      &       3.25    & 2.34   &2.24 &  2.93    &$ 0.0040$         \\
4 &    0.66    &    3.32      &      5.32       &     4.13   &  3.75   &   3.89  & 2.28  &$0.0054$     \\
5 &    0.83    &     3.46     &       5.66 &       4.14    &   3.31   &   3.54 & 2.87  & $0.0048 $      \\
7 &    1.06    &     4.12     &       6.12  &      4.31     &   4.43   &     5.75 &  3.5   &0    \\
			\hline\hline
		\end{tabular}
	\end{center}
\label{t:bigger_d}
 \end{table}

 \begin{table*}[!ht]
\caption{Higher spectral dimensions: the exponents for the  quadratic terms, and the specific heat exponent $\alpha$.}  
\begin{center}
\begin{tabular}[c]{c c c c c c c c c c c c }
\hline\hline
\quad $ p$\quad\quad & \quad\quad$\phi^\Omega_{2,1;0} $\quad \quad& \quad\quad$\phi^\Omega_{2,1;1} $\quad\quad& \quad\quad  $\phi^\Omega_{2,2} $ \quad\quad&  \quad\quad $ \phi^\Omega_{2,2^\prime}$ \quad\quad & \quad  \quad$\phi^\Omega_{2,2^{\prime\prime}}$\quad\quad
&\quad $r^\ast\quad$ & $Y_t$\quad& \quad$\alpha$\quad\quad & \quad$\delta$\quad\quad \\
\hline
2& $0.83\pm 0.06$        &      $0.14\pm 0.04$      &$0.02\pm 0.08$&$-0.04\pm 0.03$&$0.10\pm 0.03$  &0&$1$ &  0& $\infty$  \\
3&  $1.27\pm0.08 $&   $0.77\pm0.07$  &    $1.06\pm 0.08$& $0.73\pm 0.13$&$0.90\pm 0.06$&$-0.113096
$&1&1.14& 6\\
4&$1.82\pm0.24 $&$1.49\pm0.22$  &   $1.91\pm 0.26$& $1.37\pm 0.21$&$1.49\pm 0.09$&$ -0.06192$&1&0.58& 4.03\\
5&$1.58\pm0.12$&$ 1.36\pm0.12$ &$1.69\pm 0.13$& $1.31\pm 0.11$&$1.34\pm 0.16$&$-0.1490197
$&1&0.75& 3.41 \\
7&$1.91\pm0.24$&$ 1.78\pm0.23 $&$2.03\pm 0.25$& $1.71\pm 0.23$&$1.76\pm 0.23$&$0$&1&0 &3 \\
			\hline\hline
		\end{tabular}
	\end{center}
\label{t:quadratic_bigger_d}
 \end{table*}

We now consider the iterative equation for $r$, which is in the form $r^\prime = a_r r-c_r$, where the coefficients depend upon $v^\ast$.
Defining the scaling exponents $\phi_{2,1;0}$ and $\phi_{2,1;1}$  for the zeroth and first order terms of the expansion in $r_0$ of $\sigma^{-1}_{2,1}$,  we have 
\be
a_r=B^{p_2} - 12 v^\ast B^{\frac{1}{2}\zeta -\phi_{2,1;1}}\;\;,
\label{eq:a_r}
\ee
and 
\begin{multline}
c_r= 12 v^\ast B^{\frac{1}{2}\zeta}[B^{-\phi_{2,1;0}}\; \\ - 6 v^\ast (B^{-\phi_{2,2}} + B^{-\phi_{2,2^\prime}}+\tfrac{2}{3}B^{-\phi_{2,2^{\prime\prime}}})]\;\;.
\label{eq:c_r}
\end{multline}  
The fixed point $r^\ast$ is  unstable, with $ \vert a_r \vert >1$ and negative since  the critical temperature increases with  the interactions.

The linearized transformation matrix for $r$ and $v$ is upper triangular, and   $\partial r^\prime/\partial r\vert_{v^\ast}=a_r \sim B^{Y_t}$ by our definition.  
The critical exponent $\alpha$ can now be found from differentiating Eq.~(\ref{eq:fOmega}) twice with respect to $t$ and using $c_h\sim t^{-\alpha}$.  One gets $\alpha=2-p_1/Y_t$.  We calculate  $Y_t$  from Eq.~(\ref{eq:a_r}) by comparing the exponents of the first and second terms in $a_r$ , and  we find  $\zeta/2-\phi_{2,1;1} > 1$  all cases where $v^\ast\neq 0$. 

For $p=7$,  with $v^\ast=0$, the only surviving term in $a_r$ is the Gaussian one.  Then $Y_t=1$ and  from Eq.~(\ref{eq:alfa_p}) we find $\alpha=2-(1+\beta)$.  However,   for  $\beta=1.06>1$,  the ultraviolet singularity in
\be
c_h \sim \int_0^\Omega \frac{d\om \;\om^\beta}{(r_0+\om)^2} \;\;\label{eq:uv}
\ee
 takes over and we find $\alpha=0$. The values of $\alpha$ for different $p$ are given in Table~\ref{t:quadratic_bigger_d}.

Since the field renormalization $z$ has not changed in this second order perturbation expansion, the eigenvalue for the external field $h$, namely  $B^{Y_h}=z$ is still given by Eq.~(\ref{eq:delta_p}). In terms of $\beta$ we have  $\delta= (2+\beta)/\beta$.  The values are listed in  Table~\ref{t:quadratic_bigger_d}. Clearly for $\beta=1$, we get $\delta=3$, the mean field value.  Actually one may show by the methods of Section IIIC that for $\beta>1$ (or $\sd>4$), where $v_0$ is an irrelevant variable,  that $\delta=3$.

\section{Conclusions and discussion}

The eigenvalue spectrum of the graph Laplacian has already been used to define an effective dimensionality which governs both vibrational  and diffusive behavior, as well as the infrared singularities of the Gaussian model~\cite{Dhar,Cassi,Hattori} on arbitrary lattices.  Bradde {\it et al.}~\cite{Bianconi2010} have discussed the Ginzburg criterion for phase transitions on {\it spatial} complex networks in terms of the spectral dimension ${\tilde d}$ of the closely related adjacency matrix.
The present paper is complementary to the approach of Bradde et al.~\cite{Bianconi2010}, in the sense that we study non-spatial networks,  although both the Cayley tree and the diamond lattice are embeddable in two dimensions. 

We build  the spectral renormalization group method on the generalized Fourier transform using the eigenvectors of the graph Laplacian. We show how to implement a renormalization group a l\`a Wilson  on non-translationally invariant non-spatial networks, either by changing the upper cutoff of the eigenvalues of the graph Laplacian (in place of the momentum cutoff on Bravais lattices) or by scaling the total number of nodes (which is the same as the number of eigenvectors of the graph Laplacian). The lack of translational invariance forces us to avoid length-rescaling altogether, since the only thermodynamically meaningful length, the correlation length, is not well defined on these lattices~\cite{GriffithsandKaufman}. The hyperscaling relations cannot be used, since it is not quite clear how the dimension should be defined~\cite{GriffithsandKaufman} (embedding dimension $d$, fractal dimension $d_f$, or spectral dimension $\sd$). It should be noted that the {\it exact} real space renormalization group for Ising spins on the diamond lattice, applied blindly as a length rescaling, yields an eigenvalue for the reduced  temperature   $y_t=0.7479$, a correlation length exponent $\nu=1.3370$ and, if the hyperscaling relation (with $d=d_f=\sd=2$) is used, a rather embarrasing  $\alpha = -0.6738$.

For the Gaussian model, we find that the critical exponents depend exclusively on the spectral dimension through $\beta$, the scaling exponent of the spectral density in the infrared region.  Note that  $\beta=\tfrac{\sd}{2}-1$. The Gaussian fixed point is stable with respect to the introduction of quartic couplings in $\sd=2$. This  seems to be due to the fact that the spectral dimensionality $\sd$ is equal to the lower critical value within the Ising universality class,  rather  than to  the lack of translational invariance~\cite{Melrose} or some other exotic property of the diamond lattice.  It has been remarked by Wilson and Kogut~\cite{Wilson3}( p. 124),  that  reaching the non-trivial fixed point is difficult in two (Euclidean) dimensions, as the fixed point functions  acquire a sensitive dependence on the choice of the initial function to be iterated. Several authors~\cite{Bagnuls,Morris} have found that in fact, scalar field theories in two dimensions may not be amenable to perturbative RG schemes.  

We have gone to higher fractal and spectral dimensions by using a generalized diamond lattice~\cite{Itzykson}, using an integer parameter, $p$ (see Fig.~\ref{fig:pb_lattice}).  The spectral dimension is raised from two towards four  ($\beta=1$),  we encounter stable non-Gaussian fixed points.  We are able to calculate the eigenvalues for all the different diagrams in the perturbation expansion up to second order, and compute the  specific heat exponent.  As the spectral dimension is raised beyond four, although the fixed point still exists, it looses its stability. (In fact iterations for the coupling constant $v_0$ exhibit a period four attractor, however this does not carry much physical meaning because it depends purely on the truncation of the perturbation expansion).

Note that, due to the strong dependence on the eigenvectors of the graph Laplacian that is introduced by the 4-vertex $\Phi$ (Eq.~\ref{eq:Phi}), we are forced to resort to numerical computations in order to extract the eigenvalues, and beyond Gaussian theory we are not able to express them  solely  in terms of  the spectral dimension.  However, better analytical understanding of the symmetries of the eigenvectors should allow us to extract more information.  Work is under progress in this direction. 

We conclude that critical fluctuations on non-translationally invariant networks are amenable to investigation by the present method.  Only near lower critical dimensions for the universality class under study, the perturbative scheme might break down.  

One  question is how to go to higher orders in perturbation theory.  In this paper we have gone only up to second order in the coupling constant and/or the reduced temperature.  This, however, is not enough to introduce corrections to the field renormalization $z$, and the computation of the critical exponent $\eta$ may be problematic for the same reasons as that of $\nu$. 

{\bf Acknowledgements}

It is a pleasure to thank E. Brezin, B. Derrida, J.-P. Eckmann, M. Mungan, H. \"Ozbek and T. Turgut for a number of interesting discussions.  A. Tuncer acknowledges partial support from the ITUBAP No. 36259.  
\appendix
\setcounter{figure}{0} \renewcommand{\thefigure}{A.\arabic{figure}}
\setcounter{table}{0} \renewcommand{\thetable}{A.\arabic{table}}
\setcounter{section}{0} \renewcommand{\thesection}{A.\arabic{section}}
 \setcounter{equation}{0} \renewcommand{\theequation}{A.\arabic{equation}} 
\section{Square and Cubic Lattices}
\begin{figure}[!ht]
\vspace{1pt} 
\includegraphics[width=0.45\textwidth]{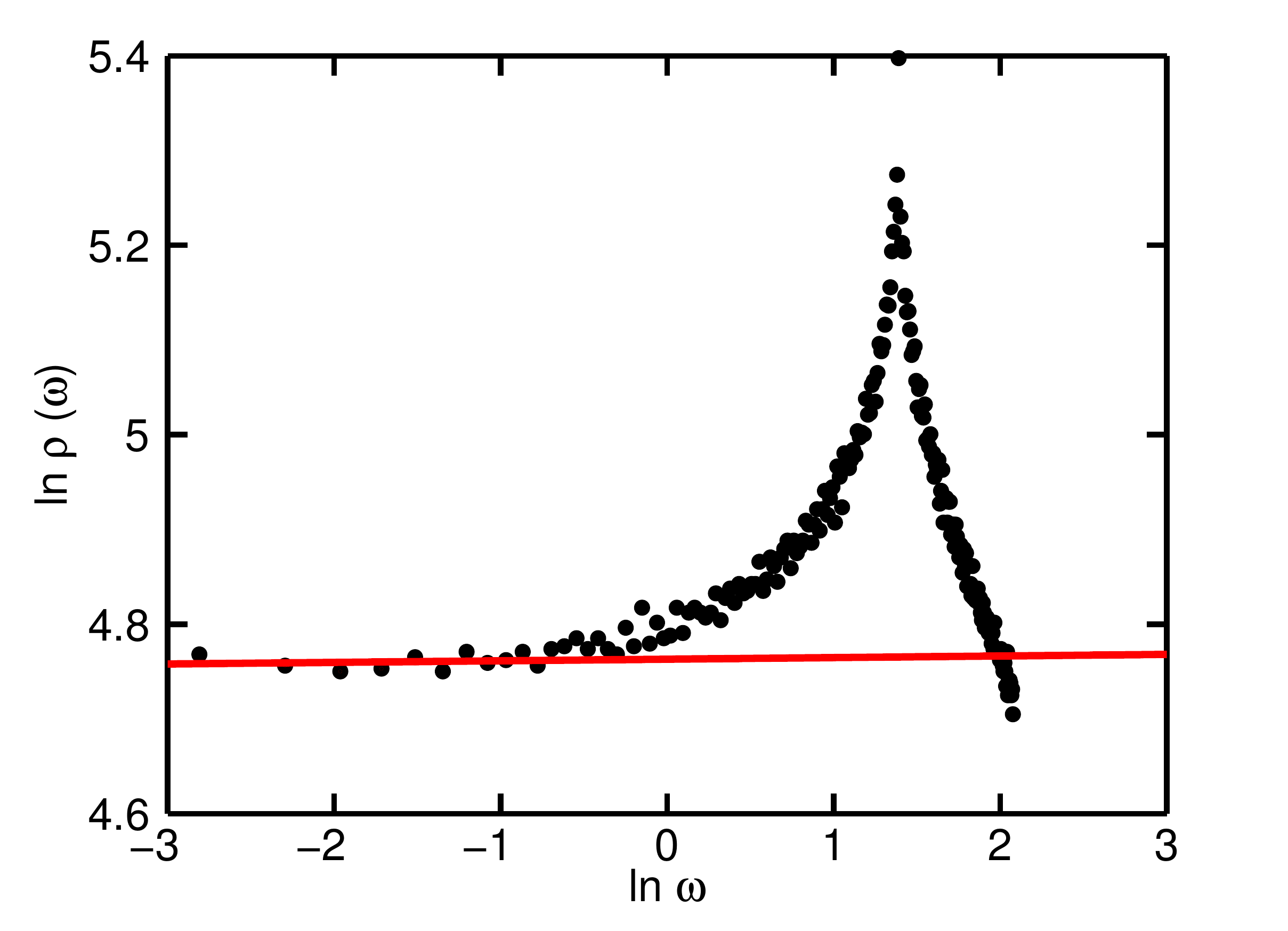}
\caption{The spectral density for the 200$\times$ 200 square lattice, $\rho(\omega)\sim \omega^\beta$, with $\beta=0.00\pm0.02$ for small $\omega$.  The  first 13 points have been fitted. }\label{fig:square_rho} 
\end{figure}

\begin{figure}[!ht]
\vspace{1pt} 
\includegraphics[width=0.45\textwidth]{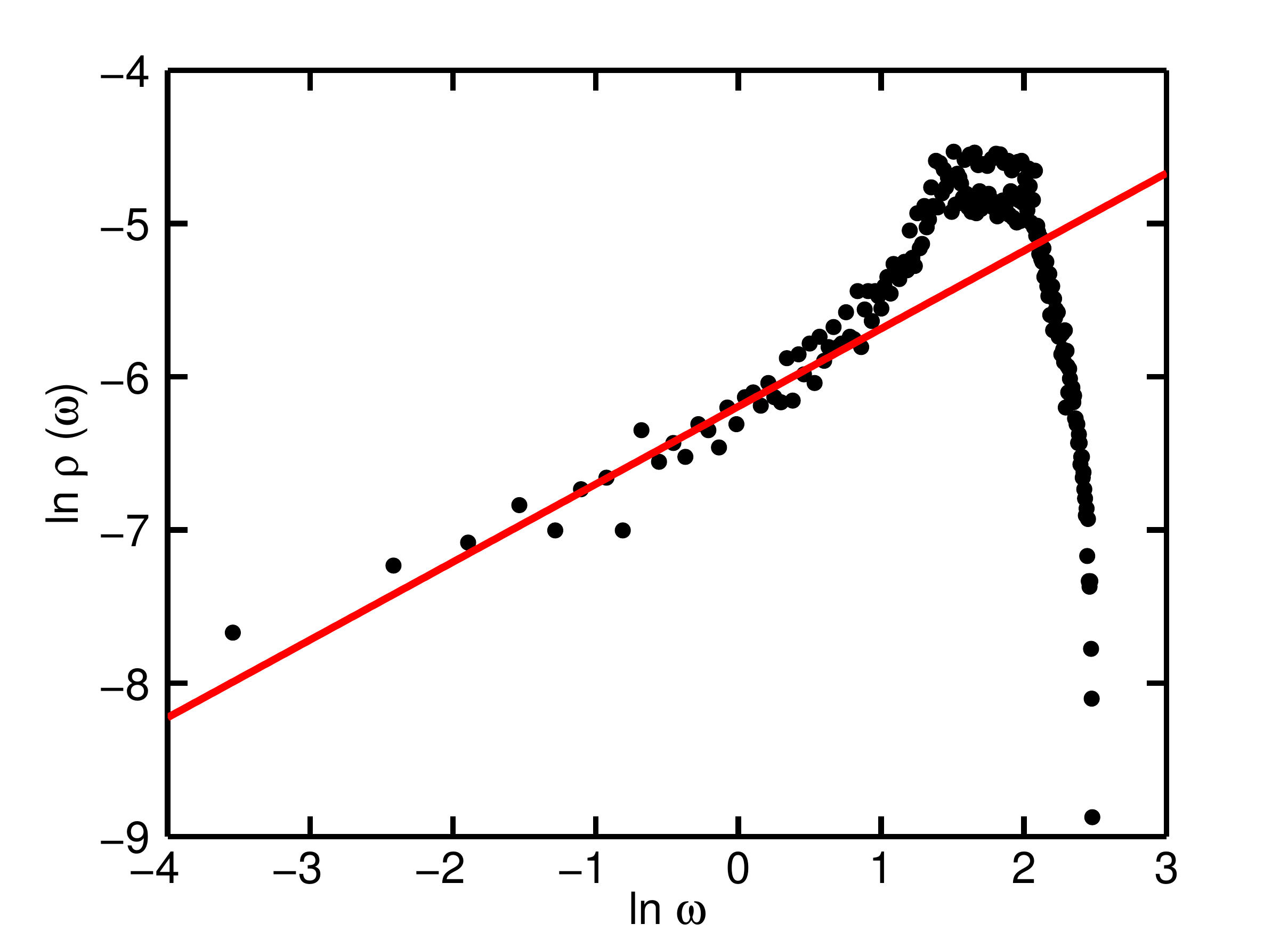}
\caption{The spectral density  for the $N=35^3=42875$ cubic lattice with periodic boundary conditions. $\rho(\omega)\sim \omega^\beta$, with $\beta=0.50 \pm 0.03 $.  Here $\rho(\omega)$ is fitted from the  2nd to the 35th point. }\label{fig:cubic_rho} 
\end{figure}

\begin{figure}[!ht]
\vspace{1pt} 
\includegraphics[width=0.45\textwidth]{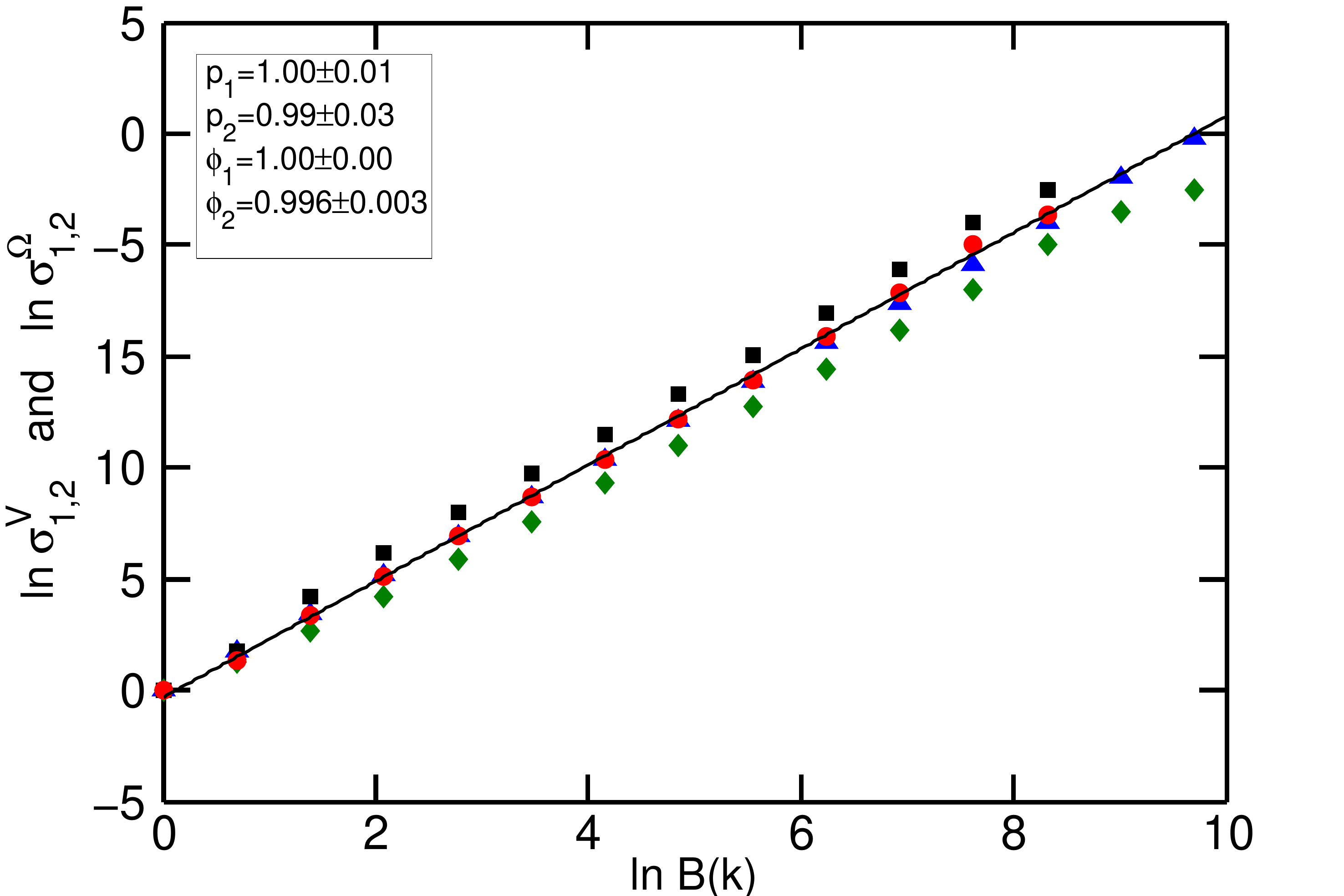}
\caption{(Color online)The rescaling factors $\sigma^{\rm V}_1$ (blue, triangles),  $\sigma^{\rm V}_2$ (black, squares), $\sigma^{\Omega}_1$ (red, circles)  and $\sigma^{\Omega}_2$ (green, diamonds) for the square lattice, some symbols fall on top of each other.  The exponents are $\phi_1, \phi_2, p_1, \; p_2$  are given in the legend. The scale factors are $B(k)=N/k$,  $k=1,...,N/2$. The scaling region is taken from $k=4$ to $k=1000$ .}\label{fig:squarefii} 
\end{figure}

\begin{figure}[!ht]
\vspace{1pt} 
\includegraphics[width=0.45\textwidth]{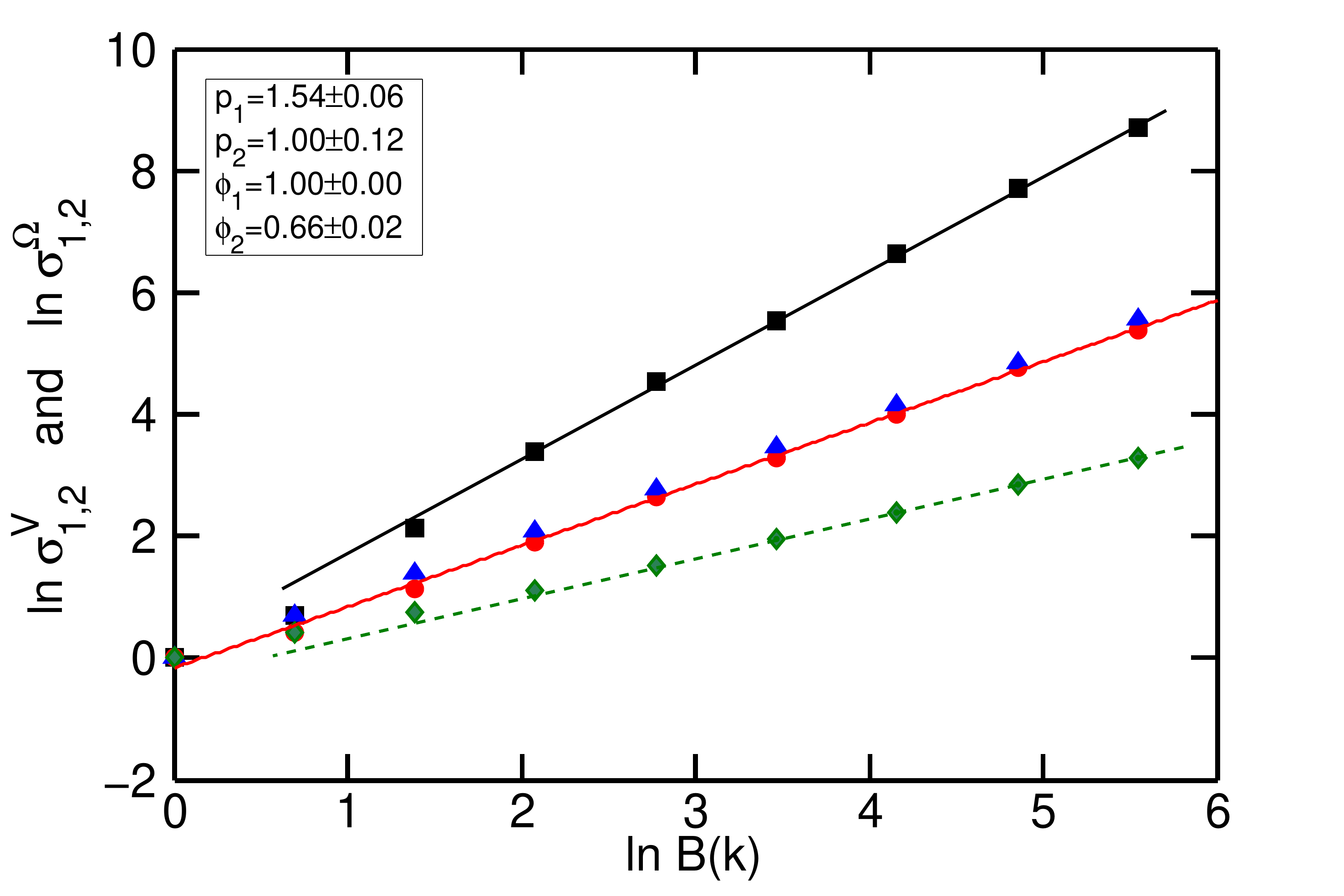}
\caption{(Color online) The RG factors $\sigma^{\rm V}_1$ (blue, triangles), $\sigma^{\rm V}_2$ (black, squares), $\sigma^{\Omega}_1$ (red, circles) and $\sigma^{\Omega}_2$ (green, diamonds) for the cubic lattice with periodic boundary conditions. All the points after the first two have been fitted.  The numerical values of the exponents  are provided in the legend.}\label{fig:cubicfii} 
\end{figure}

In Figs.~\ref{fig:square_rho},\ref{fig:cubic_rho}, we present plots for the numerical calculation of the spectral density and the exponent $\beta$  for the square lattice ($N=4\times 10^4$) and the cubic lattice  ($N=(35)^3$).  In Figs.~\ref{fig:squarefii}, \ref{fig:cubicfii}, the scaling plots for  the exponents $\phi_1, \;\; \phi_2,\;\;p_1$ and $p_2$ are displayed. A summary of all the critical exponents computed in this paper  was given in Table~\ref{allexponents} in the main text.

For many networks, even  in the thermodynamic limit the spectral density does not  become a smooth function which behaves as a power law, $\rho(\om) \sim \om^\beta$ for small $\om$. In particular, the spectral density of the  Barabasi-Albert network~\cite{BA}, as well as scale free networks with $\gamma>3$  grow exponentially for small $\om$ (so that the spectral dimension diverges), although they have a power law tail for large $ \om$, Ref.~\cite{erzan}.) As in the case of the Cayley tree and the diamond hierarchical lattice, the spectral density may be nonzero only on a union of discrete sets of measure zero. 
Where one cannot compute $\rho(\om)$ analytically, one should be aware that its numerical determination may call for prohibitively large network sizes, as has also been noted by Bradde {\it et al.}~\cite{Bianconi2010}.  Even for Bravais lattices with p.b.c., the convergence of the numerical spectral density to the thermodynamic limit is very slow, especially in the small $\om$ region, see Figs.~\ref{fig:square_rho},\ref{fig:cubic_rho}.

 \begin{table*}[!ht]
\caption{The eigenvalues and eigenvectors of the Cayley tree with two generations ($r=2$) and branching number  $b=3$.  The integers  in the first line of the table below indicate the $nth$ distinct eigenvalue $\om^{(n)}$.The eigenvalues are, in increasing order,  $\om^{(1)}=0, \;\om^{(2)}=(\alpha_+ -1)/\alpha_+\;, \om^{(3)} =1, \om^{(4)}=b+1-\sqrt{b}, \;\om^{(5)}=(\alpha_- -1)/\alpha_- = 1/\om^{(2)} , \; \om^{(6)} = b+1+\sqrt{b} $, with $\alpha_\pm = \frac{1\pm \sqrt{1+4/b}}{2}$.  Defining $\theta=2\pi/b$, the numerical factors appearing  in the eigenvectors $\mu= 2,3,11,12$ and $4,\ldots,9$ are the real and imaginary parts of $\exp(i \ell\theta), \; \ell=1,2,\ldots,b$.  The constants appearing in columns marked 4 and 6 are $p=-b-\sqrt{b}$, and $q= - b+\sqrt{b}$. The normalization constants have not been included for clarity.  Notice the wide disparity between the $2$nd and $5$th, and resp. $4$tn and $6$th eigenvalues belonging to eigenvectors with the same symmetries. 
}
\begin{center}
\begin{tabular}[c]{c c c c c c c c c c c c c  }
\hline\hline\\
1&2&2&3&3&3&3&3&3&4&5&5&6\\
\hline\\
1 & 0  	&0  & 0                & 0 &0&                        0                      &	0                                          &0&$-\sqrt{b^3}$&0   &0&  $+\sqrt{b^3}$ \\
  1 & -1/2  & $\sqrt{3}/2$  	&  0&  0 &0&			0                     &  0                                          &0  &p &-1/2& $\sqrt{3}/2$ 	&$q$ \\  
  1 &-1/2   &$ -\sqrt{3}/2$ 	& 0 &    0   &0&         0                      &   0                                          & 0   &p& -1/2     & 	 $-\sqrt{3}/2$ & $q$\\
  1 &  1 	&   0		&   0   & 0 &0 &                    0                     &  0			                              &0 	&p		&	1&   0&$ q$\\
  1& $-(1/2)  \alpha_+$ &  $(\sqrt{3}/2)  \alpha_+ $&     -1/2& 0 &0& $\sqrt{3}/2$ &0                       &0& 1&     $-(1/2)  \alpha_-$                & $(\sqrt{3}/2)  \alpha_- $&  1\\
 1& $ -(1/2)  \alpha_+$ 	&$(\sqrt{3}/2)  \alpha_+$  & -1/2 &0  &0&  -$\sqrt{3}/2$   &0                   &0&1&                   $ -(1/2)  \alpha_-$ &$(\sqrt{3}/2)  \alpha_- $&   1\\
 1& $ -(1/2)  \alpha_+	$ &$(\sqrt{3}/2)  \alpha_+ $ &       1& 0 &0& 0  &  0                                       &0&1&$ -(1/2)  \alpha_-$ &$(\sqrt{3}/2)  \alpha_- $&   1\\
 1& $ -(1/2) \alpha_+ $ 	&$- (\sqrt{3}/2)  \alpha_+$ &  0& -1/2 &  0&0&  $\sqrt{3}/2$                                      &0&1& $ -(1/2) \alpha_-$ &	$-(\sqrt{3}/2)  \alpha_- $& 1  \\
 1&$  -(1/2)  \alpha_+	$  	&$ - (\sqrt{3}/2)  \alpha_+$ & 0 & -1/2 & 0&0&    -$\sqrt{3}/2$                 &0 &1&	$ -(1/2)  \alpha_-$		&$-(\sqrt{3}/2)  \alpha_- $& 1  \\
 1&$  -(1/2)  \alpha_+	 $ 	&$ - (\sqrt{3}/2)  \alpha_+$ & 0 & 1 &     0&0&	0		                              &0  &1&$ -(1/2)  \alpha_-$ &	$-(\sqrt{3}/2)  \alpha_- $&   1\\
 1&  $  \alpha_+	$         &  0                                  &  0   &   0    &-1/2	  &0  &0     & $\sqrt{3}/2$		&1&	$   \alpha_-$	&0&1   \\
 1& $ \alpha_+ $	            &  0                                 &   0  &0        &-1/2 &0&0	&	$-\sqrt{3}/2$	     &1&	$   \alpha_-$	&0&1   \\
 1& $ \alpha_+ $	           & 0                                     & 0     & 0    &1        &   0  &0						        &0 &1&		$   \alpha_-$	&0&1   \\
		\hline\hline
		\end{tabular}
	\end{center}
\label{t:eigenvectors}
 \end{table*}


\section{Eigenvalues and eigenvectors of the graph Laplacian}

Closed form expressions for the eigenvalues of the graph Laplacian of the Cayley tree are known~\cite{Rojo,Rabbiano}; however, to our knowledge, explicit solutions for all the eigenvalues for arbitrary $r$ are not available. (Also see  Rozikov and coworkers~\cite{Rozikov2006,Rozikov2013,Rozikov_Rev_2013}).
We present in Table~\ref{t:eigenvectors},  the eigenvalues and  eigenvectors  of the graph Laplacian for the Cayley tree with r=2. 

Ochab et al.\cite{Burda} provide valuable insights  into the successive stages of symmetry breaking leading to the different eigenvectors and the  eigenvalues of the adjacency matrix of the Cayley tree. We have been able to analytically compute the eigenvectors for arbitrary $r$, for $\omega < \om^\ast$, using the automorphism properties of the Cayley tree; this will be presented in a separate publication. For larger trees, the same pattern (Table~\ref{t:eigenvectors}) is stretched to the whole tree, and then, for  increasing $\om$,  becomes localized on subtrees of diminishing size.

\begin{figure}[ht]
\vspace{1pt} 
\centering 
\includegraphics[width=0.45\textwidth]{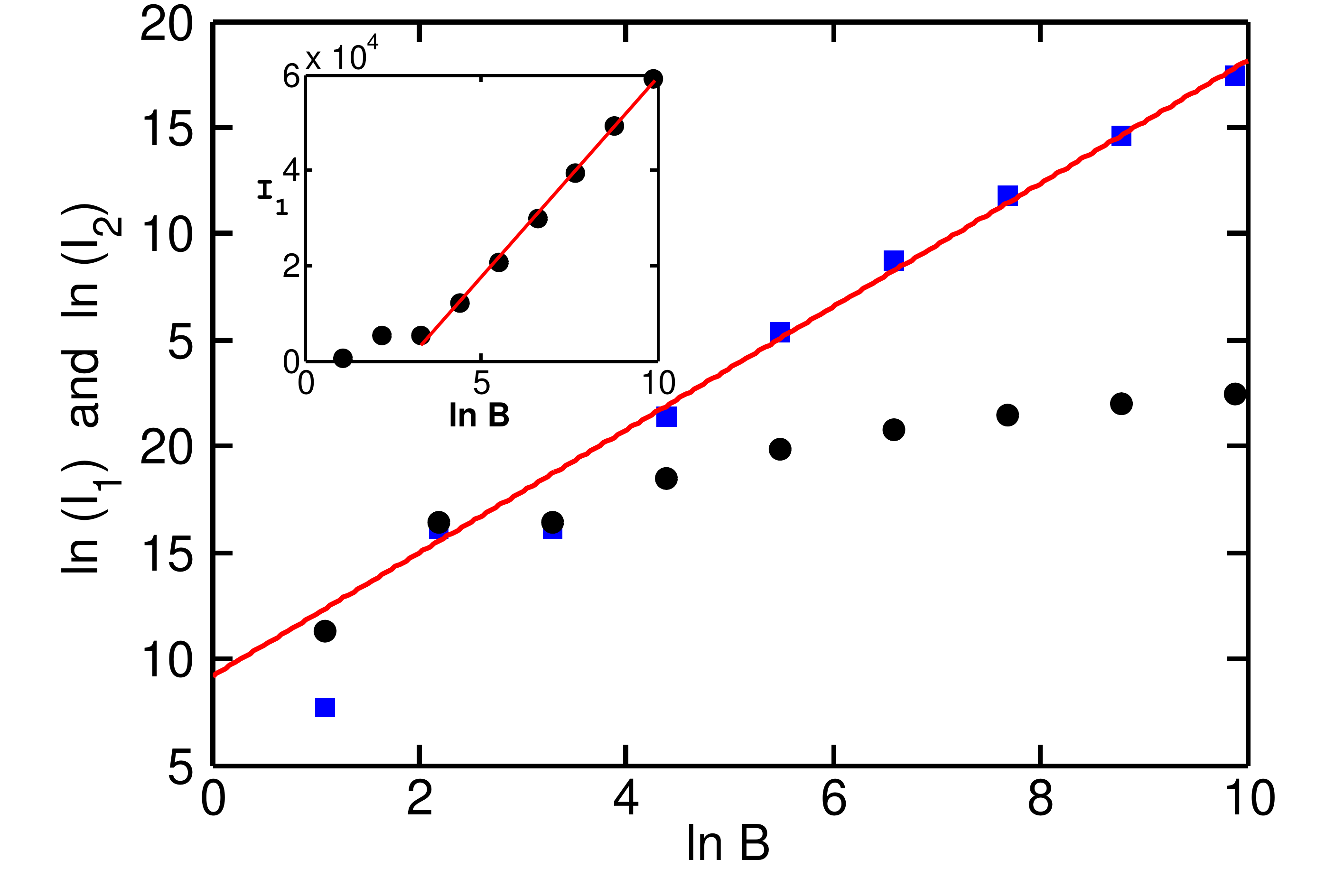}
\caption{Scaling behavior of $I_1^{(0)}$ (circles) and $I_2^{(0)}$ (squares) for the Cayley tree.  The data points for $I_1^{(0)}$  are better fitted by $\ln B$, as seen in the inset.  The slope of the fitted line is $1.09\pm0.04$, gives $I_2^{(0)} \sim B^{1.09\pm0.04}$.}  \label{fig:I1I2}
\end{figure}
%

\section {Scaling of the functions $I_K(0)$ on the Cayley tree}
Here we would like to calculate explicitly the functions $ I_K(B)$, $K=1,2$,
\be I_K(B) \equiv \sum_{\mu=\mu_B}^N \om_\mu^{-K}\;\;,\label{I_K}
\ee
 for the Cayley tree, making use of the approximate formulae (Eqs.~\ref{tau},\ref{omega}). 
Let us take the scaling factor to equal powers of the branching number $b$, so that $B_k=b^k$.  We see that already for $k=1$, the eigenvalues included in the sum will be below the peak at $\omega^\ast=1$, where the equations cited above hold to a good approximation.  Defining $n_k$ as the label of the smallest distinct eigenvalue compatible with $\mu_{B_k}$,
\be
I_K^{(0)}=A_K+\sum_{n=n_k}^{r} \tau_n [\om^{(n)}]^{-K}\;\;.\label{I_K_Cayley}
\ee
Note that for $k=r$, $N/B_r \simeq 1$.  The constant $A_K$ is the sum of $[\om^{(n)}]^{-K}$ over all $n$ such that $\om^{(n)} >1$.
Then 
\be 
I_K^{(0)}=A_K+\sum_{n=n_k}^{r}\frac {b^{(n-2)}(b-1)}{a_n b^{-(r-n+2)K}}\;\;.
\ee

Approximating $a_n$ by a constant as before, canceling $b^{n}$ in the summand  and using $n_k=r-k+1$ and $r-n_k-1 = k$ we find\ for $K=1$,  
\be
I_1^{(0)} = A_1+ \frac {(b-1) b^r}{a} k\;\;, 
\ee
or
\be
\label{I_1_Cayley}
I_1^{(0)} = {\rm const.} + \frac {(b-1) b^r}{a \ln b} \ln B_k\;\;, 
\ee
where the dependence on the scaling parameter in only logarithmic. Similarly, one gets,
\be
I_2^{(0)}=A_2+\frac{b^{r+2}(b^k-1)}{a^2} \ee
giving
\be
I_2^{(0)}={\rm const.} + \frac{b^{r+2}}{a^2} B_k\;\;, \label{I_2_Cayley}
\ee
where the dependence on the scaling parameter $B$ is linear.  It should be noted that $b^r \propto N$ in all of the above.

\end{document}